\documentclass[pra,twocolumn,superscriptaddress]{revtex4-2}

\usepackage{amsmath}
\usepackage{amsfonts}
\usepackage{graphicx}
\usepackage{siunitx}
\usepackage{soul}
\usepackage[colorlinks=true,linkcolor=blue]{hyperref}

\usepackage{soul}
\usepackage{xcolor}

\begin{document}

%\title{\tab{Jaz} Husimi-driven many-body systems realized with Bose-Einstein condensates}
\title{Atomtronic Many-Body Transport using Husimi Driving}
%\title{Atomtronic BEC Transport using Husimi-Driving}
%\title{An Husimi Based Atomtronic Transport Methods}

% author list
\author{B.J.~{Mommers}}
\affiliation{Centre for Engineered Quantum Systems, The University of Queensland, St Lucia, Australia}
\affiliation{School of Mathematics and Physics, The University of Queensland, St Lucia, Australia}
\email[]{b.mommers@uq.edu.au}
\author{A.~{Pritchard}}
\affiliation{Centre for Engineered Quantum Systems, The University of Queensland, St Lucia, Australia}
\affiliation{School of Mathematics and Physics, The University of Queensland, St Lucia, Australia}
\author{T.A.~{Bell}}
\affiliation{Centre for Engineered Quantum Systems, The University of Queensland, St Lucia, Australia}
\affiliation{School of Mathematics and Physics, The University of Queensland, St Lucia, Australia}
\author{R.N.~{Kohn, Jr.}}
\affiliation{Space Dynamics Laboratory, Albuquerque, New Mexico 87106, USA}
\author{S.E.~{Olson}}
\affiliation{Air Force Research Laboratory, Kirtland AFB, New Mexico 87117, USA}
\author{M.~{Baker}}
\affiliation{Quantum Technologies, Cyber and Electronic Warfare Division, Defence Science and Technology Group, Brisbane, Australia}
\affiliation{School of Mathematics and Physics, The University of Queensland, St Lucia, Australia}
\affiliation{Centre for Engineered Quantum Systems, The University of Queensland, St Lucia, Australia}
\author{M.W.J.~{Bromley}}
\affiliation{School of Sciences, University of Southern Queensland, Toowoomba, Australia}
\affiliation{School of Mathematics and Physics, The University of Queensland, St Lucia, Australia}

%%%%%%%%%%%%%%%%%%%%%%%%%%%%%%%%%%%%%%%%%%%%%%%
%%%%%%%%%%%%%%%%%%%%%%%%%%%%%%%%%%%%%%%%%%%%%%%
%%%%%%%%%%%%%%%%%%%%%%%%%%%%%%%%%%%%%%%%%%%%%%%
%%%%%%%%%%%%%%%%%%%%%%%%%%%%%%%%%%%%%%%% Abstract %%
%%%%%%%%%%%%%%%%%%%%%%%%%%%%%%%%%%%%%%%%%%%%%%%
%%%%%%%%%%%%%%%%%%%%%%%%%%%%%%%%%%%%%%%%%%%%%%%
%%%%%%%%%%%%%%%%%%%%%%%%%%%%%%%%%%%%%%%%%%%%%%%
\begin{abstract}
Quantum systems with exact analytic solutions are rare - challenging the realisation of excitation-free transport methods for many-body systems.
Husimi's 1953 treatment of linearly driven harmonic oscillators constitutes an important exception, describing a wavepacket which is spatially translated but otherwise unperturbed by the driving. 
In this work, we experimentally demonstrate the application of Husimi's solution to an interacting many-body system, namely optically- and magnetically-trapped Bose-Einstein condensates subject to resonant and off-resonant linear magnetic driving potentials.
The observed centre-of-mass motion is consistent with theory and shows minimal excitation of the displaced condensate -- a highly desirable property of any condensate manipulation technique.
We demonstrate transport 72 times faster than adiabatic rates, and a novel Husimi driving-based trap frequency measurement.
%TAB: the word "some" seems too casual - remove word?
%We finally propose some applications based on our experimental results: an atom interferometry scheme, and methods for extended transport and precision control of one-body, few-body, and many-body systems via Husimi driving.
We finally propose future applications based on our experimental results: an atom interferometry scheme, and methods for extended transport and precision control of one-body, few-body, and many-body systems via Husimi driving.
\end{abstract}

\maketitle

Expeditious and excitation-free transport methods for many-body systems have been long-sought for quantum control~\cite{2019gueryodelin0a}, having applications in quantum information, quantum simulation, and many developing atomtronics applications~\cite{2021amico0a}. Most demonstrated approaches have achieved transport through the translation of potentials with fixed spatial profiles. These trajectories have been optimised using invariant-based methods~\cite{2011chen0a,2011torrontegui0a,2012torrontegui0a,2012torrontegui0b,2018chen0a}, reverse engineering~\cite{2015zhang0a}, and control theory~\cite{2007hohenester0a,2014gueryodelin0a,2014deffner0a,2016zhang0a,2019amri0a} -- under the field of shortcuts to adiabaticity (STA). Implementations for cold atom transport include magnetic conveyors~\cite{2001greiner0a,2001hansel0b,2014minniberger0a,2019badr0a,2013han0a,2010chen0a}, optical conveyors ~\cite{2018ness0a,couvert08a,2001gustavson0a,leonard14a,2012middelmann0a}, and optical lattices~\cite{2006schmid0a}. 

Experimental implementations capable of fast Bose-Einstein condensate (BEC) transport have received comparatively little attention. BEC transport using optical lattice~\cite{2003mandel0a,2005browaeys0a} and tweezer~\cite{2001gustavson0a} methods has been demonstrated -- though unwanted excitations remained. While proposed STA methods promise improvements in BEC fast transport~\cite{2019amri0a,2018corgier0a,2012torrontegui0a,2007hohenester0a}, much of this theoretical work remains untested, and some STA designs have been found unsuitable for experimental use~\cite{2018ness0a}. Analytical solutions which enable the centre-of-mass transport of many-body interacting systems without intrinsic heating have not been demonstrated.

%Few quantum many-body systems are exactly solvable. 
We leverage and expand Husimi's seminal 1953 solution for a single-particle confined in a forced quantum harmonic oscillator~\cite{1953husimi0a} to demonstrate the fast and excitation-free transport of $^{87}\text{Rb}$ BECs using two complementary experimental implementations. Husimi considered two scenarios: an amplitude modulated Harmonic potential, and the combination of a harmonic and an amplitude-modulated linear spatial potential. The first scenario attracted the most attention~\cite{1994kleber0a,1996kagan0a}, since it describes systems with time-dependent trapping frequency variations~\cite{1950feynman0a,1960dykhne0a,1969solimeno0a,1969popov0a,2009lohe0a}. The second remained a novel analytic solution ~\cite{1951feynman0a,1952fujiwara0a,1958kerner0a,1994kleber0a}.

%%%%%%%%%%%%%%%%%%%%%%%%%%%%%%%%%%%%%%%%%%%%%%%%%%%%%%%%
\begin{figure}[!t]
\includegraphics[width=0.9\linewidth]{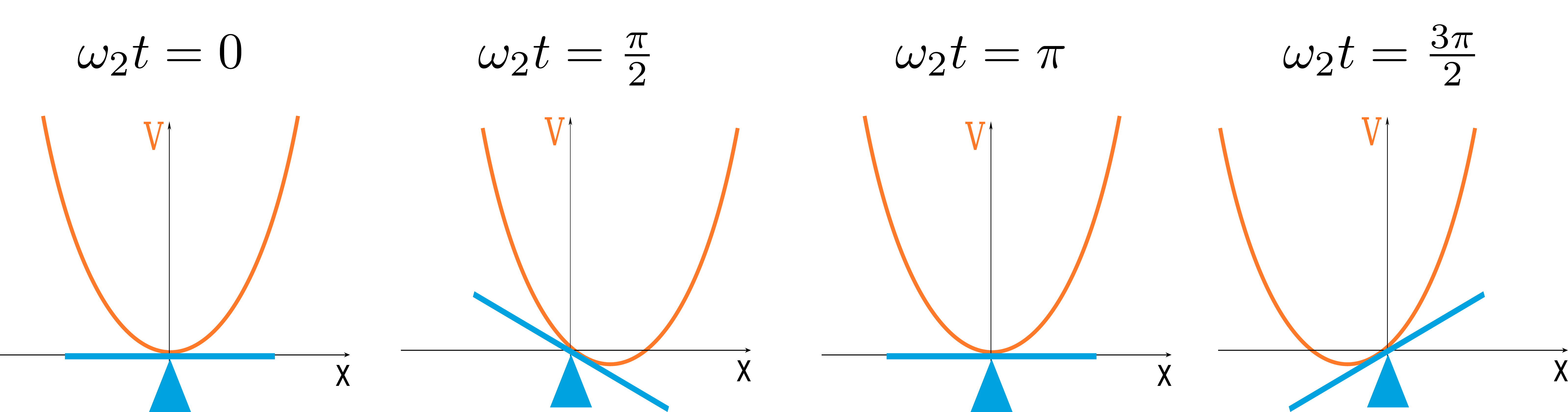}
\includegraphics[width=0.9\linewidth]{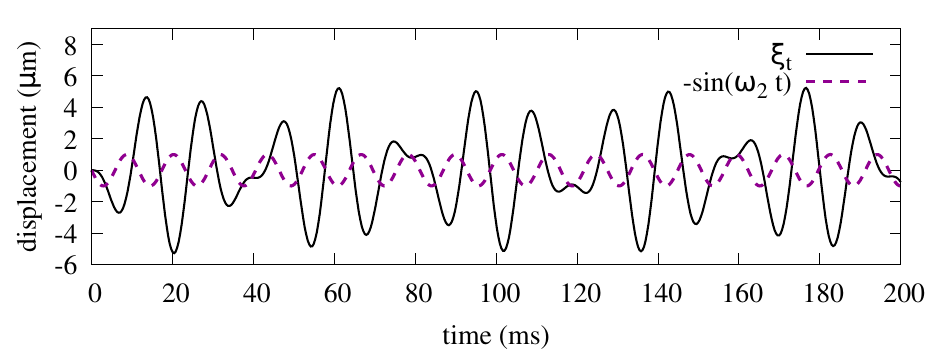}
\caption[Standard Husimi]{\label{fig:husimi-standard}
(color online) Top: Schematic illustration of Husimi driving. The combination of a
fixed harmonic trap with an oscillating linear potential,
$V(x,t) = \frac12 \omega_1^2 x^2 - x \sin(\omega_2 t)$,
can be visualized as a harmonic trap balancing on an oscillating see-saw -- distinct from lateral displacements of the trap.
Bottom: Analytic centre-of-mass displacement $\xi_t$ (black solid) with the driving function $-\sin(\omega_2 t)$ (purple dashed).
The frequency ratio $\omega_2/\omega_1 = \sqrt{2}$ results in aperiodic motion of the centre-of-mass~\cite{2021jackson0a} -- a ratio chosen for ease of experimental comparison. The corresponding field gradient $b_x = 0.85$G/cm.}
\label{fig:concept}
\end{figure}
%%%%%%%%%%%%%%%%%%%%%%%%%%%%%%%%%%%%%%%%%%%%%%%%%%%%

We address the second scenario in the context of many-body quantum systems, see~Fig.~\ref{fig:concept}. Herein called Husimi driving, the single-particle solution exhibits centre-of-mass oscillations which preserve the arbitrary initial state~\cite{1953husimi0a,1991bandrauk0a,1994kleber0a,1998hanggi0a,1998dittrich0a,2021jackson0a}. Introducing an amplitude-modulated linear potential is shown to identically displace the center-of-mass for all spatial modes of the initially harmonic potential - the probability density distribution about the centre-of-mass is unaffected by the Husimi driving-induced centre-of-mass motion. 

The single particle solution persists when introducing particle interactions - complete derivations are described in the supplementary material~\cite{supplementary1}.
We show the results are unaffected by arbitrary scattering interactions using both the mean-field Gross-Pitaevskii equation (GPE) and the more rigorous $N$-particle Hamiltonian, findings consistent with the Harmonic Potential Theorem (HPT) for many-body systems~\cite{1994dobson0a}.
%TAB: suggested reword?
%The absence of driving-induced relative motion between modes in the many-body solution ensures our Husimi transport protocol avoids additional thermal friction effects at finite temperature.
%\tab{The absence of driving-induced relative motion between modes in the many-body solution ensures our Husimi transport protocol overcomes the thermal friction effects which would otherwise induce damping at finite temperature.}
The absence of driving-induced relative motion between all modes in the many-body solution predicts that under Husimi driving, finite-temperature thermal friction effects are avoided.

% potential equation broken out per readability guidelines for PRA
The relevant 1D Husimi driving potential consists of static harmonic and amplitude-modulated linear terms, 
\begin{equation}
V(x,t) = \frac12 m\hspace{0.2ex}\omega_1^2 x^2 - A \hspace{0.2ex} x f(t),
\end{equation}
where $A$ represents the modulation amplitude, $f(t)$ the time-dependent modulation function, $m$ the atomic mass, and $\omega_1$ the angular trapping frequency along the driving dimension $x$ within a Cartesian coordinate system. This potential is distinguishable from rigid translations given it also energetically decreases and increases (per Fig. \ref{fig:concept}). The distinction from rigid translation is best shown by rewriting the driving potential,
\begin{equation}
   V(x,t) \equiv \frac12 m \hspace{0.2ex}\omega_1^2 \hspace{0.1ex} \left(x - \frac{A \hspace{0.1ex}f\hspace{-0.2ex}(t)}{m\hspace{0.2ex}\omega_1^2}\right)^2
       - \frac{A^2 f^2\hspace{-0.2ex}(t)}{2m\hspace{0.2ex}\omega_1^2}~.
       \label{eqn:ModTrap}
\end{equation}

%%%%%%%%%%%%%%%%%
% Continue editing from here...

%As Husimi driving within an ideal 3-D Harmonic trapping configuration will not excite additional modes beyond those present in the underlying NLSE dynamics $\chi(x-\xi,y,z,t)$~\cite{supplementary1}, these results also hold for the full 3-D NLSE. 

Our foundational theoretical introduction of particle interactions directly extends Husimi's original treatment of the Schr\"odinger equation to the Gross-Pitaevskii equation (GPE),
\begin{equation}
i \hspace{0.05ex}\hbar \hspace{0.05ex}\frac{\partial\hspace{0.05ex}\psi}{\partial \hspace{0.05ex}t} = \left( \! -\frac{\hbar^2}{2\hspace{0.05ex}m} \frac{\partial^2}{\partial x^2} + V(x,t) + g \hspace{0.05ex} |\psi|^2 \! \right) \! \psi ~,
\label{eqn:GPE}
\end{equation}
which describes the evolution of the condensate order parameter $\psi(x,t)$. The contact interaction parameter $g$ quantifies the $s$-wave atom-atom scattering strength~\cite{1997esry0a}. The GPE is a nonlinear Schr{\"o}dinger equation describing low-temperature condensate dynamics. We solve for the order parameter $\psi(x,t)$ using $f(t) = \sin(\omega_2 t)$, and the coordinate transformation $x^\prime = x - \xi(t)$. This solution is independent of the interaction term in Eqn.~\ref{eqn:GPE}, and results in a time-dependent centre-of-mass oscillation $\xi(t) = \langle x \rangle$. 
As with Husimi's original solution, $f(t)$ may take an arbitrary form, and our choice of function is motivated by experimental convenience.
In the co-moving coordinate of $\xi(t)$, the GPE solution reduces to that for a static harmonic trap $\chi(x,t)$ with a time-varying phase factor,
\begin{equation}
\psi(x,t) = \chi(x-\xi,t) \hspace{0.2ex} \exp \hspace{-0.7ex} \left[ i \hspace{0.2ex} m \hspace{0.2ex} \dot{\xi} (x-\xi)/\hbar + i E \hspace{0.1ex} t/\hbar \right] ~.
\label{eqn:husimi-soln}
\end{equation}

BEC spatial density distributions $n(x,t)=\left|\psi(x,t)\right|^2$ are unaffected by the global phase changes and therefore preserved under Husimi driving. The initial state instead evolves according to $\chi(x-\xi,t)$, about the center-of-mass $\xi$. The centre-of-mass oscillations $\xi(t)$ are simply solutions to the classical forced oscillator. Under resonant driving $\omega=\omega_1=\omega_2$,
\begin{align}
\xi_0(t) &= \frac{A}{2m\hspace{0.1ex} \omega^2}
           \left[ \hspace{0.2ex} \sin\left(\omega\hspace{0.1ex} t\right) - \omega \hspace{0.1ex}t \cos(\omega \hspace{0.1ex}t)\hspace{0.1ex}\right]~,\vspace{20ex}
\label{eqn:xiRes}
\end{align}
and non-resonant driving $\omega_1\neq\omega_2$,
\begin{align}
\hspace{-1ex}\xi_1(t)\hspace{-0.2ex} &=\hspace{-0.2ex} \frac{A}{m \hspace{-0.2ex}\left(\omega_1^2\hspace{-0.4ex} - \hspace{-0.4ex} \omega_2^2\right)} \hspace{-0.5ex}
           \left[ \hspace{0.1ex} \sin\left(\hspace{-0.2ex}\omega_2 t\right)\hspace{-0.2ex} - \hspace{-0.2ex} \frac{\omega_2}{\omega_1}\hspace{0.1ex} \sin(\omega_1 t) \exp(\gamma t) \right] \hspace{-0.2ex}.
\label{eqn:xiNon}
\end{align}
The non-resonant driving expression for $\xi_1(t)$ features an experimentally-motivated damping factor $\gamma$ -- which we later describe.
%We additionally introduced an experimentally-motivated phenomenological damping factor $\gamma$ into the expression for $\xi_1(t)$ -- motivation is subsequently provided.

%%%%%%%%%%%%%%%%%%%%%%%%%%%%%%%%%%%%%%%%%%%%%%%
\begin{figure*}[!t]
\includegraphics[width=\linewidth]{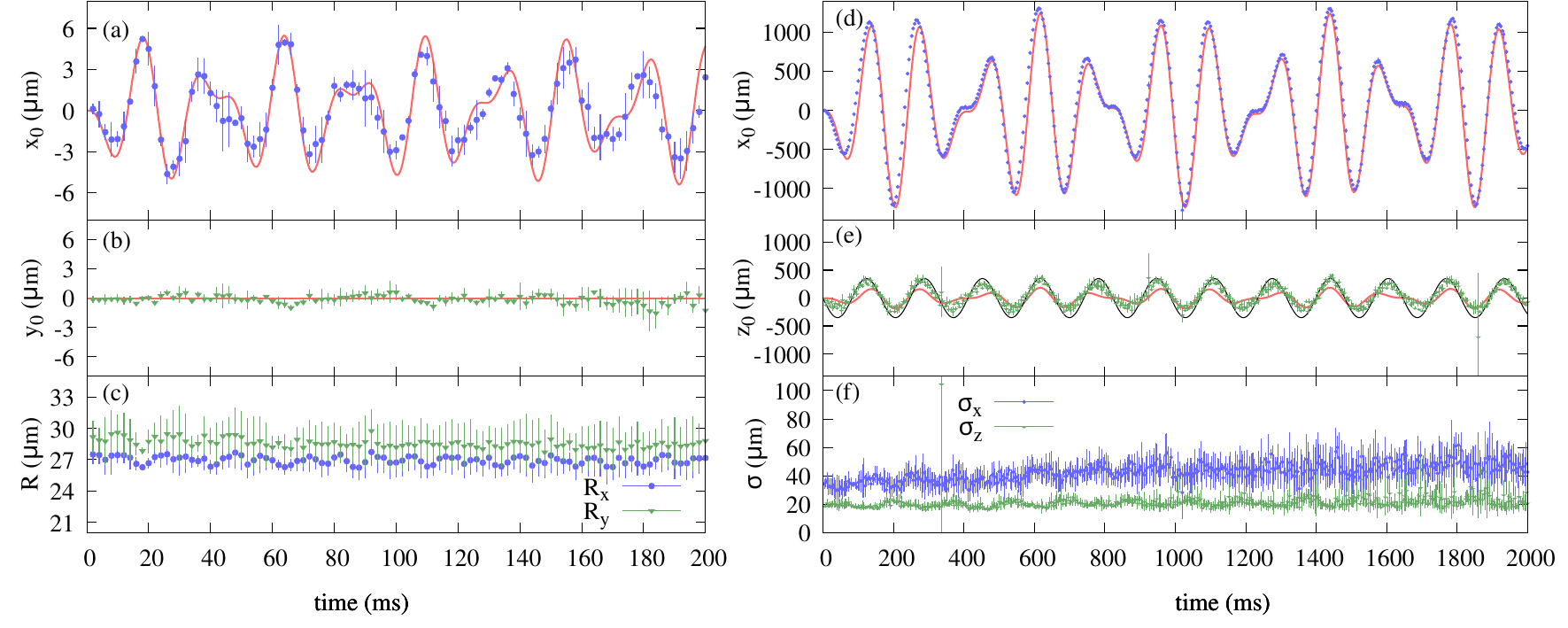}
\caption[Standard Husimi]{(color online)
Aperiodic experimental Husimi data associated with the non-resonant driving frequency $\omega_2/\omega_1=1/\sqrt{2}$. 
%%%%%%%%%%%% OPTICAL DIPOLE %%%%%%%%%%%%%%
(a-c)~The optical dipole experiment data. Each data point is the average of three runs, with the error bars indicating one standard deviation.
Panels (a) and (b) show the centre-of-mass along the driven $x$- and orthogonal $y$-direction, respectively.
The solid (red) lines show the fits to Eqn.~\ref{eqn:xiNon} with parameters: $\omega_1 = 2\pi (66.4 \pm 0.2)$~Hz,
$\omega_2 = 2\pi \cdot 43.13$~Hz (fixed),
$b_x = 0.93 \pm 0.03$~G/cm, and
$\gamma = 6.7 \pm 1.1$~s$^{-1}$.
(c)~The Thomas-Fermi radii $R_x$ and $R_y$ along the driven and orthogonal image axes, respectively.
%%%%%%%%%%%% ATOM-CHIP %%%%%%%%%%%%%%
(d-f)~Equivalent data for the atom-chip experiment. Each data point is the average of four runs, with one standard deviation indicated by the error bars. 
The solid (red) lines shows the fits to Eqn.~\ref{eqn:xiRes}.
The regression parameters in (f) were
$\omega_1= 2\pi (8.52 \pm 0.01)$~Hz,
$\omega_2 = 2\pi ( 6.08 \pm 0.01)$~Hz,
$b_x = 3.17 \pm 0.06$~G/cm, and
$\gamma = 0.00 \pm 0.02$~s$^{-1}$;
while in (e) were
$\omega_1= 2\pi (29.98 \pm 0.01)$~Hz,
$\omega_2 = 2\pi ( 6.066 \pm 0.003)$~Hz,
$b_z = 23.8 \pm 0.6$~G/cm, and
$\gamma = 0$.
(f)~The $1/e$ Gaussian radii $\sigma_x$ and $\sigma_y$ along the driven and orthogonal axes, respectively.
The raw data appears in Supplementary Fig.~5~\cite{supplementary1}.
}
\label{fig:HusimiAperiodic}
\end{figure*}
%%%%%%%%%%%%%%%%%%%%%%%%%%%%%%%%%%%%%%%%%%%%%%%%%%%%%%%%

We establish the viability and versatility of Husimi driving for BEC manipulation and control using two distinct experimental trapping methods. BECs are formed using optical-dipole \cite{2016bell0a,2018bell0a} and magnetic-dipole \cite{2017stickney0a, 2016squires0a} harmonic potentials, and the $\left|F,m_F\right>=\left|1,-1\right>$ and $\left|2,2\right>$ spin-states, respectively.
Larger mass densities achieved using optical trapping enhance the particle interactions and motivate a bimodal Thomas-Fermi plus Gaussian fit for the finite-temperature spatial density distribution.
Larger condensate fractions achieved using magnetic confinement motivate a unimodal Gaussian fit for the density distribution.
Collective excitations induced through Husimi driving would be reflected by oscillations in the condensate Thomas-Fermi radii $R_x$ and $R_y$, or Gaussian waists $\sigma_x$ and $\sigma_y$, respectively.
Heating induced through Husimi transport would deplete the BEC, reducing the Thomas-Fermi radii but broadening the Gaussian waists over time, respectively. 
Integrated experimental density distributions were measured through imaging along the orthogonal $z$ axis.

Our optical potential was formed using orthogonal and focused $1064$~nm Gaussian beams. An elliptical sheet beam confined against gravity; the $1/e^2$ waists $1.25$~mm (in $x$) and $27$~$\mu$m (in $z$) gave the trapping frequency $\omega_z \approx 2\pi \cdot 140$~Hz. Cylindrical confinement across the sheet was achieved using a pinning beam with waists $89.7$~$\mu$m (in $x$) and $87.6$~$\mu$m (in $y$). The trapping frequency $\omega_1\approx 2\pi (61\pm3)$~Hz was initially measured using the parametric heating method. 
A linear magnetic potential was formed using approximately Helmholtz configured coils positioned symmetrically about the condensate, which produced the potential $V_B(x,t) = \mu_B g_F m_F b_x x \sin(t)$; where $\mu_B$ is the Bohr magneton, $g_F$ the Land{\'e} factor, and $b_x$ the field amplitude gradient (in x). Condensates were formed with atom numbers between $N=(7.2\pm0.4)\times10^4$ and $(3.4\pm0.1)\times10^5$, and condensate fractions between $N_0/N=0.47\pm0.05$ and $0.69\pm0.08$. Absorption images were captured from above using $20$~ms time-of-flight.

%%%%%%%%%%%%%%

Our second apparatus used independently-controlled atom-chip wires to produce configurable magnetic traps $V(x,t) \approx c_2 x^2 + c_1(t) x$, where we choose to suppress higher-order terms. This generates both the static harmonic term and oscillating linear potential terms required for Husimi driving.  The trapping frequency $\omega_1 = 2\pi  (8.545\pm0.003)$~Hz was measured by rapidly turning on $c_1(t)$ from zero to a fixed value to shift the location of the harmonic potential minimum and observing the subsequent coherent state oscillation. Condensates with approximately $10^4$ atoms and a high condensate fraction were imaged from the side using $8$~ms time-of-flight.

Non-resonant Husimi oscillations were driven using the optical experiment with the frequency $\omega_2 = 2 \pi \cdot 43.13$~Hz, roughly corresponding to $\omega_2/\omega_1\approx1/\sqrt{2}$.  The resulting aperiodic centre-of-mass trajectories are shown in Fig.~\ref{fig:HusimiAperiodic}(a-c). Three data sets were recorded sequentially to evaluate consistency. No appreciable coupling of the centre-of-mass motion along the driven $x$ and orthogonal $y$ spatial directions was observed. The approximately constant Thomas-Fermi radii additional indicates no appreciable heating or collective mode excitation.

Husimi oscillations were driven using the atom-chip experiment with frequency $\omega_2 = 2\pi \cdot 6.042$~Hz, which again corresponds to $\omega_2/\omega_1\approx1/\sqrt{2}$. The corresponding centre-of-mass trajectories are shown in Fig.~\ref{fig:HusimiAperiodic}(d-f). Four data sets were recorded sequentially to evaluate consistency; the error bars shown are mostly imperceptible. Outlying data which resulted from sporadic issues during the condensate formation and imaging processing has been retained in Fig.~\ref{fig:HusimiAperiodic}. The cloud width along the $x$-direction shows a growth trend, and minimal change in $y$, possibly indicating some heating.

Our numerical regressions for the experimental data in Fig.~\ref{fig:HusimiAperiodic}(a,d) using Eqn.~\ref{eqn:xiNon} validate the solution derived for non-resonant driving. However, the optical experiment Husimi oscillations in Fig.~\ref{fig:HusimiAperiodic}(a) were observed to decay over longer periods of driving, with the center-of-mass eventually following the linear field (see Supplementary Figure~1~\cite{supplementary1}). This observation motivated our inclusion of the phenomenological damping $\gamma$ in Eqn.~\ref{eqn:xiNon}, which we numerically determined to be $\gamma = -5.4 \pm 0.4$~s$^{-1}$ for our optical trapping configuration.

\begin{figure}[!t]
\includegraphics[width=\linewidth]{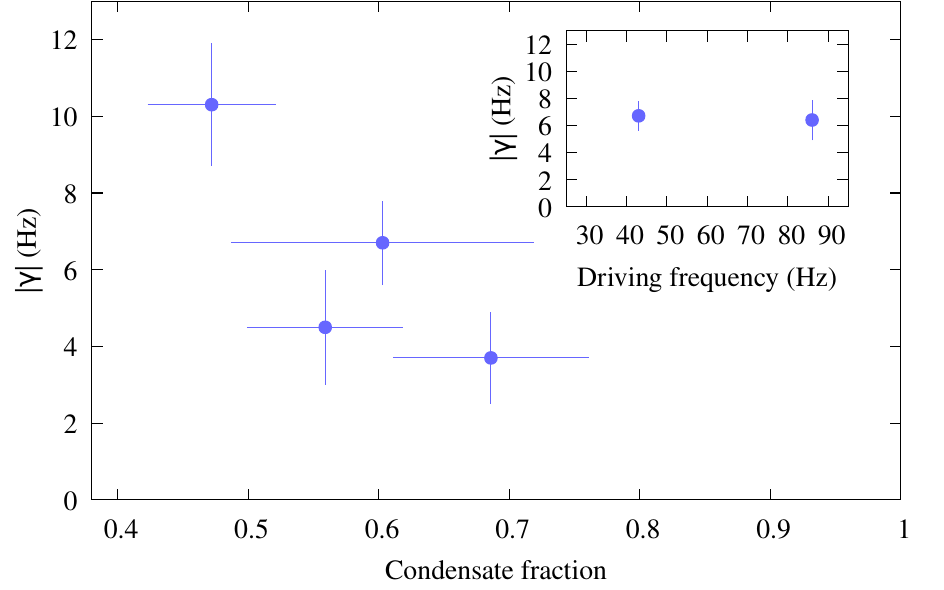}
\caption[Standard Husimi]{(color online)
Experimentally-observed damping in the optical dipole experiment
under Husimi with trap frequency of $\omega_1 \approx 2\pi \cdot 61\pm3$~Hz.
The damping rate $\gamma$ is extracted from the fit to the experimental
data using Eqn.~\ref{eqn:xiNon}. 
The main figure shows the result of varied condensate fraction $N_0/N$
with a fixed driving frequency of $\omega_2 = 2\pi \cdot 43.13$~Hz.
The inset shows an additional data point by driving at $\omega_2 = 2\pi \cdot 86.27$~Hz
(also giving anharmonic motion) with a constant condensate fraction ($\approx 0.6$).
}
\label{fig:fig03-damping}
\end{figure}

We investigated the damping systematics by varying the condensate fraction and driving frequency. Fig.~\ref{fig:fig03-damping} shows the fitted values of the damping parameter, which we found to be independent of driving frequency, but dependent on condensate fraction. We attribute this damping to anharmonicity of the optical potential - only harmonic traps produce a Husimi-like solution with identical displacement of all spatial modes. Imperfections across the Gaussian beam and the Gaussian form of the trap itself are sources of anharmonicity that could not be fully eliminated and may contribute to the observed trajectory damping. Our 2D simulations mimic the experiment and support this conclusion - see Supplementary Section 3~\cite{supplementary1}. A lower condensate fraction indicates a larger thermal cloud, and atoms can potentially scatter between condensate modes and the thermal fraction due to the anharmonicity. The absence of damping in the atom chip experiment ($\gamma \approx 0)$ suggests the anharmonic $x^3, x^4, \hdots$ trapping terms have been  minimised -- even when accounting for the expected longer damping time scale caused by the lower trap and drive frequencies used in the atom chip experiment.

Whilst the atom-chip experiment suffered negligible damping, appreciable center-of-mass oscillations along the transverse $y$-direction are evident in Fig.~\ref{fig:HusimiAperiodic}(e). This non-resonant Husimi trajectory was the result of strong synchronous driving along $y$, caused by our experimental control limitations. A regression of the transverse oscillation data found $b_y/b_x \approx 7.5$, $\omega_y = 2\pi (29.98 \pm 0.01)$~Hz, and $\omega_2 = 2\pi (6.066 \pm 0.003)$~Hz. The faster and smaller oscillation features, while difficult to discern, allowed us to determine $\omega_y$ with a reduced uncertainty.

%%%%%%%%%%%%%%%%%%%%%%%%%%%%%%%%%%%%%%%%%%%%%%%%%%%%%%%%
%%%%%%%%%%%%%%%%%%%%%%%%%%%%%%%%%%%%%%%%%%%%%%%%%%%%%%%%

We next consider applications of Husimi driving as a platform for high-precision BEC experiments.
Transport is achieved by `catching' the condensate at the stationary points $\dot{\xi}(t) = 0$.  These points occur at times where $f(t)=0$ when driving on resonance. By temporarily disabling the linear potential and displacing the remaining harmonic potential to be centred on the condensate's centre of mass, it is possible to catch the stationary atoms at the new position. Repetition of this process can allow for long-distance excitation-free transport.  

\begin{figure}[b]
\includegraphics[width=\linewidth]{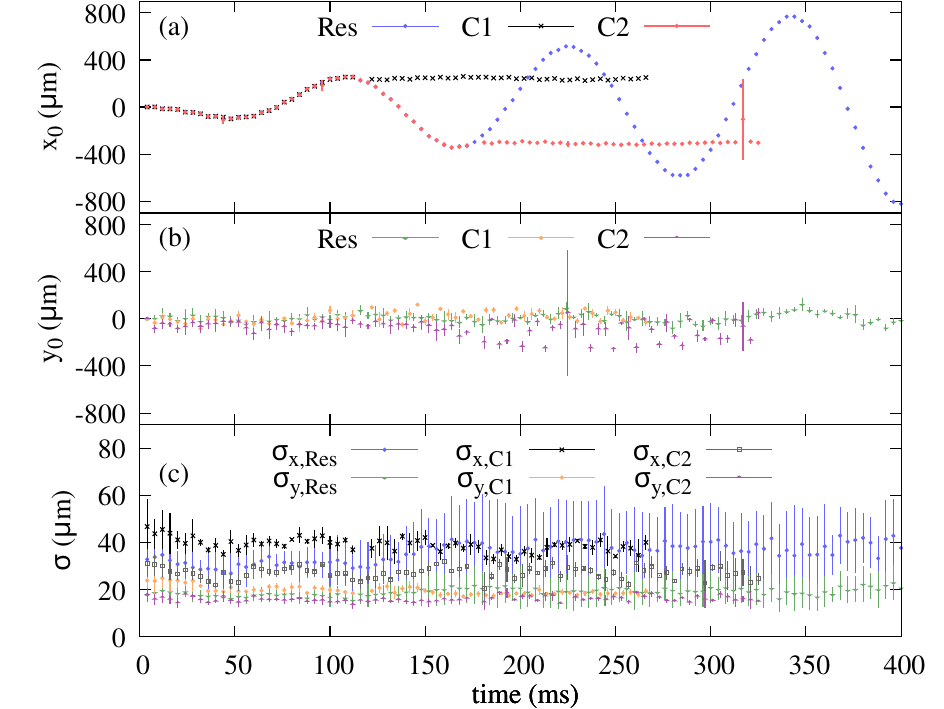}
\caption[Standard Husimi]{(color online)
Atom chip experiment for Husimi driven transport
with trap frequency of $\omega_1 \approx 2\pi \cdot 8.545\pm0.003$ Hz
and resonant driving frequency set to $\omega_2 = 2\pi \cdot 8.545$~Hz
Three experiments are shown --- `Res' is the resonant driving data runs
that follows Eqn.~\ref{eqn:xiRes},  whilst `C1' and `C2' are
the two transport runs that catch the atoms at the turning points.
Panels (a),(b) show centre-of-mass measurements in the $x$- and $y$-direction, respectively.
Panel (c) shows the cloud width ($\sigma$) measured in
the $x$- and $y$-directions, with marker types and colours matching (a) and (b).
The raw data for these experiments appears in Supplementary Fig.~6~\cite{supplementary1}.
}
\label{fig:exp04-afrltransport}
\end{figure}

Our proof-of-concept resonant Husimi driving based transport scheme was demonstrated with the atom chip experiment, as shown in Fig.~\ref{fig:exp04-afrltransport}(a) using  $\omega_1 = 2\pi( 8.545\pm0.003)$ Hz and $\omega_2 = 2\pi \cdot 8.545$~Hz. By rocking the atoms backwards then forwards, the resonant motion predicted by Eqn.~\ref{eqn:xiRes} is observed, including the linear increase in the oscillation amplitude. Longer time data is shown in Supplementary Fig.~7~\cite{supplementary1}. Atoms were captured at the second and third turning points by adjusting the trapping parameter $c_1(t)$. The horizontal data post transport exhibits very little residual oscillation.

%%%%%%%%%%%%%%%%%%%%%%%%%%%%%%%%%%%%%%%%%%%%%%%%%%%%%%%%
%%%%%%%%%%%%%%%%%%%%%%%%%%%%%%%%%%%%%%%%%%%%%%%%%%%%%%%%

The stability of the condensate after the harmonic trap displacement indicates that repetition of the transport procedure could be repeated from the new centre-of-mass position. Repeated shifts then allow conveyor belt transport while maintaining the initial BEC state. Near-resonant Husimi data from the optical dipole experiment appears in the Supplementary Fig.~4~\cite{supplementary1}). Future implementations can achieve transport by rapidly translating the pinning laser using an acousto-optic deflector or spatial light modulator.

To compare transport speeds obtained via different experimental techniques, an implementation-agnostic measure of transport speed is required. We consider an effective oscillator speed as a measure of the rate of transport.
% Replace "constrains" with "characterises"?
This effective speed constrains how rapidly a trap can be displaced without inducing excitations. $^{87}$Rb atoms in an $\omega=2\pi\cdot8.545$~Hz trap have an oscillator length $\beta = \sqrt{\hbar/m \omega} \approx 3.7$~$\mu$m, oscillator period $T = 2\pi/\omega \approx 120$~ms, and therefore an oscillator speed $\beta/T \approx 30$~$\mu$m/s $=0.03$~$\mu$m/ms. We define the transport speedup factor as the ratio of the average transport speed to the oscillator speed, allowing for comparisons between different transport protocols, trapping methods, and atomic species.

Applying this measure, the Husimi-based transport in Fig.~\ref{fig:exp04-afrltransport} has a speedup factor of $72$ (data series `C1'). This is modest in relation to STA methods which have proposed/theoretical speedups of up to $1000$~\cite{2012torrontegui0a} -- though experimental work has shown additional constraints are required to reduce excitation, resulting in implementation of transport with speedups in the range of $200$ to $500$~\cite{2018corgier0a}. However, our results do not explore the limit of Husimi transport speed. Faster transport is achievable by increasing the driving gradient $b_x$ and driving frequency $\omega_2$ in Eqn.~\ref{eqn:xiRes}. The centre-of-mass trajectory in Fig.~\ref{fig:exp04-afrltransport} already indicates that a speedup of $270$ is achievable in the existing atom chip apparatus by catching the condensate at the first maximum of the trajectory.

Another application of Husimi driving is the non-destructive \textit{in-situ} measurement of the trapping frequency $\omega_1$. Our proof-of-principle regression results in Fig.~\ref{fig:HusimiAperiodic} provide a more precise measure of the optical trap frequency compared to parametric heating, and can simultaneously quantify the magnetic field gradient and the presence of anharmonicity through fitting the damping parameter. For practical use, this method must enable greater measurement accuracy without requiring greater quantities of data. We outline an approach for minimising the experimental data required in the Supplementary Section 4~\cite{supplementary1}. Other driving functions $f(t)$ in Eqn.~\ref{eqn:ModTrap} could also be exploited. 

Beyond the applications demonstrated in this work, the separability of the Husimi solution in multiple dimensions would also enable interferometry in 3-D. Lissajous-like trajectories can be engineered using unequal driving frequencies along orthogonal directions, including closed loops for Sagnac interferometry. This gives a similar scheme to that recently demonstrated by the group of Sackett \textit{et al}.~\cite{2020moan0a}, and could also be performed with condensates in separate harmonic wells.

In conclusion, our demonstrations of Husimi driving of a many-body system show that the platform is promising for the manipulation and study of BEC systems. We have shown that Husimi driving can be achieved using both existing magneto-optical and purely magnetically trapped systems, with only minimal modification. We show fast transport comparable to STA schemes is achievable and, with further optimisation, higher speeds and excitation-free transport can be realized. We establish Husimi driving provides a repeatable method for \textit{in situ} measurement of harmonic trap frequencies without deforming or displacing the trap.

\begin{acknowledgments}
The authors thank
C.D.~Jackson, O.~Sandberg, T.W.~Neely, and M.J.~Davis for valuable discussions,
and especially thank H.~Rubinsztein-Dunlop for her support.
This research was initiated through an Australian
Research Council Future Fellowship (FT100100905),
and supported by the Australian Research Council Centre of Excellence for
Engineered Quantum Systems (EQUS, CE170100009).
Numerical simulations were performed on The University of Queensland's School of Mathematics
and Physics Core Computing Facility ``\texttt{getafix}"
(thanks to Dr.~L.~Elliott and I.~Mortimer for computing support).
A portion of this work was performed in the
Space Vehicles Directorate of the US Air Force Research Laboratory
with funding from the Air Force Office of Scientific Research (AFRL/AFOSR Lab Task 19RVCOR0499).
\end{acknowledgments}

\bibliography{atom-optics-qld}

\begin{thebibliography}{51}%
\makeatletter
\providecommand \@ifxundefined [1]{%
 \@ifx{#1\undefined}
}%
\providecommand \@ifnum [1]{%
 \ifnum #1\expandafter \@firstoftwo
 \else \expandafter \@secondoftwo
 \fi
}%
\providecommand \@ifx [1]{%
 \ifx #1\expandafter \@firstoftwo
 \else \expandafter \@secondoftwo
 \fi
}%
\providecommand \natexlab [1]{#1}%
\providecommand \enquote  [1]{``#1''}%
\providecommand \bibnamefont  [1]{#1}%
\providecommand \bibfnamefont [1]{#1}%
\providecommand \citenamefont [1]{#1}%
\providecommand \href@noop [0]{\@secondoftwo}%
\providecommand \href [0]{\begingroup \@sanitize@url \@href}%
\providecommand \@href[1]{\@@startlink{#1}\@@href}%
\providecommand \@@href[1]{\endgroup#1\@@endlink}%
\providecommand \@sanitize@url [0]{\catcode `\\12\catcode `\$12\catcode
  `\&12\catcode `\#12\catcode `\^12\catcode `\_12\catcode `\%12\relax}%
\providecommand \@@startlink[1]{}%
\providecommand \@@endlink[0]{}%
\providecommand \url  [0]{\begingroup\@sanitize@url \@url }%
\providecommand \@url [1]{\endgroup\@href {#1}{\urlprefix }}%
\providecommand \urlprefix  [0]{URL }%
\providecommand \Eprint [0]{\href }%
\providecommand \doibase [0]{https://doi.org/}%
\providecommand \selectlanguage [0]{\@gobble}%
\providecommand \bibinfo  [0]{\@secondoftwo}%
\providecommand \bibfield  [0]{\@secondoftwo}%
\providecommand \translation [1]{[#1]}%
\providecommand \BibitemOpen [0]{}%
\providecommand \bibitemStop [0]{}%
\providecommand \bibitemNoStop [0]{.\EOS\space}%
\providecommand \EOS [0]{\spacefactor3000\relax}%
\providecommand \BibitemShut  [1]{\csname bibitem#1\endcsname}%
\let\auto@bib@innerbib\@empty
%</preamble>
\bibitem [{\citenamefont {{Gu{\'e}ry-Odelin}}\ \emph
  {et~al.}(2019)\citenamefont {{Gu{\'e}ry-Odelin}}, \citenamefont {Ruschhaupt},
  \citenamefont {Kiely}, \citenamefont {Torrontegui}, \citenamefont
  {{Mart{\'i}nez-Garaot}},\ and\ \citenamefont {Muga}}]{2019gueryodelin0a}%
  \BibitemOpen
  \bibfield  {author} {\bibinfo {author} {\bibfnamefont {D.}~\bibnamefont
  {{Gu{\'e}ry-Odelin}}}, \bibinfo {author} {\bibfnamefont {A.}~\bibnamefont
  {Ruschhaupt}}, \bibinfo {author} {\bibfnamefont {A.}~\bibnamefont {Kiely}},
  \bibinfo {author} {\bibfnamefont {E.}~\bibnamefont {Torrontegui}}, \bibinfo
  {author} {\bibfnamefont {S.}~\bibnamefont {{Mart{\'i}nez-Garaot}}},\ and\
  \bibinfo {author} {\bibfnamefont {J.~G.}\ \bibnamefont {Muga}},\ }\bibfield
  {title} {\bibinfo {title} {Shortcuts to adiabaticity: Concepts, methods, and
  applications},\ }\href {https://doi.org/10.1103/RevModPhys.91.045001}
  {\bibfield  {journal} {\bibinfo  {journal} {Rev.~Mod.~Phys.}\ }\textbf
  {\bibinfo {volume} {91}},\ \bibinfo {pages} {045001} (\bibinfo {year}
  {2019})}\BibitemShut {NoStop}%
\bibitem [{\citenamefont {Amico}\ \emph {et~al.}(2021)\citenamefont {Amico},
  \citenamefont {Boshier}, \citenamefont {Birkl}, \citenamefont {Minguzzi},
  \citenamefont {Miniatura}, \citenamefont {Kwek},\ and\ \citenamefont {{et
  al}}}]{2021amico0a}%
  \BibitemOpen
  \bibfield  {author} {\bibinfo {author} {\bibfnamefont {L.}~\bibnamefont
  {Amico}}, \bibinfo {author} {\bibfnamefont {M.}~\bibnamefont {Boshier}},
  \bibinfo {author} {\bibfnamefont {G.}~\bibnamefont {Birkl}}, \bibinfo
  {author} {\bibfnamefont {A.}~\bibnamefont {Minguzzi}}, \bibinfo {author}
  {\bibfnamefont {C.}~\bibnamefont {Miniatura}}, \bibinfo {author}
  {\bibfnamefont {L.}~\bibnamefont {Kwek}},\ and\ \bibinfo {author}
  {\bibnamefont {{et al}}},\ }\bibfield  {title} {\bibinfo {title} {{Roadmap on
  Atomtronics: State of the art and perspective}},\ }\href
  {https://doi.org/10.1116/5.0026178} {\bibfield  {journal} {\bibinfo
  {journal} {AVS~Quantum~Sci.}\ }\textbf {\bibinfo {volume} {3}},\ \bibinfo
  {pages} {039201} (\bibinfo {year} {2021})}\BibitemShut {NoStop}%
\bibitem [{\citenamefont {Chen}\ \emph {et~al.}(2011)\citenamefont {Chen},
  \citenamefont {White}, \citenamefont {Borries},\ and\ \citenamefont
  {{DeMarco}}}]{2011chen0a}%
  \BibitemOpen
  \bibfield  {author} {\bibinfo {author} {\bibfnamefont {D.}~\bibnamefont
  {Chen}}, \bibinfo {author} {\bibfnamefont {M.}~\bibnamefont {White}},
  \bibinfo {author} {\bibfnamefont {C.}~\bibnamefont {Borries}},\ and\ \bibinfo
  {author} {\bibfnamefont {B.}~\bibnamefont {{DeMarco}}},\ }\bibfield  {title}
  {\bibinfo {title} {{Quantum Quench of an Atomic Mott Insulator}},\ }\href
  {https://doi.org/10.1103/PhysRevLett.106.235304} {\bibfield  {journal}
  {\bibinfo  {journal} {Phys.~Rev.~Lett.}\ }\textbf {\bibinfo {volume} {106}},\
  \bibinfo {pages} {235304} (\bibinfo {year} {2011})}\BibitemShut {NoStop}%
\bibitem [{\citenamefont {Torrontegui}\ \emph {et~al.}(2011)\citenamefont
  {Torrontegui}, \citenamefont {{Ib{\'a}{\~n}ez}}, \citenamefont {Chen},
  \citenamefont {Ruschhaupt}, \citenamefont {{Gu{\'e}ry-Odelin}},\ and\
  \citenamefont {Muga}}]{2011torrontegui0a}%
  \BibitemOpen
  \bibfield  {author} {\bibinfo {author} {\bibfnamefont {E.}~\bibnamefont
  {Torrontegui}}, \bibinfo {author} {\bibfnamefont {S.}~\bibnamefont
  {{Ib{\'a}{\~n}ez}}}, \bibinfo {author} {\bibfnamefont {X.}~\bibnamefont
  {Chen}}, \bibinfo {author} {\bibfnamefont {A.}~\bibnamefont {Ruschhaupt}},
  \bibinfo {author} {\bibfnamefont {D.}~\bibnamefont {{Gu{\'e}ry-Odelin}}},\
  and\ \bibinfo {author} {\bibfnamefont {J.~G.}\ \bibnamefont {Muga}},\
  }\bibfield  {title} {\bibinfo {title} {{Fast atomic transport without
  vibrational heating}},\ }\href {https://doi.org/10.1103/PhysRevA.83.013415}
  {\bibfield  {journal} {\bibinfo  {journal} {Phys.~Rev.~A}\ }\textbf {\bibinfo
  {volume} {83}},\ \bibinfo {pages} {013415} (\bibinfo {year}
  {2011})}\BibitemShut {NoStop}%
\bibitem [{\citenamefont {Torrontegui}\ \emph
  {et~al.}(2012{\natexlab{a}})\citenamefont {Torrontegui}, \citenamefont
  {Chen}, \citenamefont {Modugno}, \citenamefont {Schmidt}, \citenamefont
  {Ruschhaupt},\ and\ \citenamefont {Muga}}]{2012torrontegui0a}%
  \BibitemOpen
  \bibfield  {author} {\bibinfo {author} {\bibfnamefont {E.}~\bibnamefont
  {Torrontegui}}, \bibinfo {author} {\bibfnamefont {X.}~\bibnamefont {Chen}},
  \bibinfo {author} {\bibfnamefont {M.}~\bibnamefont {Modugno}}, \bibinfo
  {author} {\bibfnamefont {S.}~\bibnamefont {Schmidt}}, \bibinfo {author}
  {\bibfnamefont {A.}~\bibnamefont {Ruschhaupt}},\ and\ \bibinfo {author}
  {\bibfnamefont {J.~G.}\ \bibnamefont {Muga}},\ }\bibfield  {title} {\bibinfo
  {title} {{Fast transport of Bose-Einstein condensates}},\ }\href
  {https://doi.org/10.1088/1367-2630/14/1/013031} {\bibfield  {journal}
  {\bibinfo  {journal} {New~J.~Phys.}\ }\textbf {\bibinfo {volume} {14}},\
  \bibinfo {pages} {013031} (\bibinfo {year} {2012}{\natexlab{a}})}\BibitemShut
  {NoStop}%
\bibitem [{\citenamefont {Torrontegui}\ \emph
  {et~al.}(2012{\natexlab{b}})\citenamefont {Torrontegui}, \citenamefont
  {{Mart{\'i}nez-Garaot}}, \citenamefont {Ruschhaupt},\ and\ \citenamefont
  {Muga}}]{2012torrontegui0b}%
  \BibitemOpen
  \bibfield  {author} {\bibinfo {author} {\bibfnamefont {E.}~\bibnamefont
  {Torrontegui}}, \bibinfo {author} {\bibfnamefont {S.}~\bibnamefont
  {{Mart{\'i}nez-Garaot}}}, \bibinfo {author} {\bibfnamefont {A.}~\bibnamefont
  {Ruschhaupt}},\ and\ \bibinfo {author} {\bibfnamefont {J.~G.}\ \bibnamefont
  {Muga}},\ }\bibfield  {title} {\bibinfo {title} {{Shortcuts to adiabaticity:
  Fast-forward approach}},\ }\href {https://doi.org/10.1103/PhysRevA.86.013601}
  {\bibfield  {journal} {\bibinfo  {journal} {Phys.~Rev.~A}\ }\textbf {\bibinfo
  {volume} {86}},\ \bibinfo {pages} {013601} (\bibinfo {year}
  {2012}{\natexlab{b}})}\BibitemShut {NoStop}%
\bibitem [{\citenamefont {Chen}\ \emph {et~al.}(2018)\citenamefont {Chen},
  \citenamefont {Jiang}, \citenamefont {Li}, \citenamefont {Ban},\ and\
  \citenamefont {Sherman}}]{2018chen0a}%
  \BibitemOpen
  \bibfield  {author} {\bibinfo {author} {\bibfnamefont {X.}~\bibnamefont
  {Chen}}, \bibinfo {author} {\bibfnamefont {R.}~\bibnamefont {Jiang}},
  \bibinfo {author} {\bibfnamefont {J.}~\bibnamefont {Li}}, \bibinfo {author}
  {\bibfnamefont {Y.}~\bibnamefont {Ban}},\ and\ \bibinfo {author}
  {\bibfnamefont {E.~Y.}\ \bibnamefont {Sherman}},\ }\bibfield  {title}
  {\bibinfo {title} {{Inverse engineering for fast transport and spin control
  of spin-orbit-coupled Bose-Einstein condensates in moving harmonic traps}},\
  }\href {https://doi.org/10.1103/PhysRevA.97.013631} {\bibfield  {journal}
  {\bibinfo  {journal} {Phys.~Rev.~A}\ }\textbf {\bibinfo {volume} {97}},\
  \bibinfo {pages} {013631} (\bibinfo {year} {2018})}\BibitemShut {NoStop}%
\bibitem [{\citenamefont {Zhang}\ \emph {et~al.}(2015)\citenamefont {Zhang},
  \citenamefont {Chen},\ and\ \citenamefont
  {{Gu{\'e}ry-Odelin}}}]{2015zhang0a}%
  \BibitemOpen
  \bibfield  {author} {\bibinfo {author} {\bibfnamefont {Q.}~\bibnamefont
  {Zhang}}, \bibinfo {author} {\bibfnamefont {X.}~\bibnamefont {Chen}},\ and\
  \bibinfo {author} {\bibfnamefont {D.}~\bibnamefont {{Gu{\'e}ry-Odelin}}},\
  }\bibfield  {title} {\bibinfo {title} {{Fast and optimal transport of atoms
  with nonharmonic traps}},\ }\href
  {https://doi.org/10.1103/PhysRevA.92.043410} {\bibfield  {journal} {\bibinfo
  {journal} {Phys.~Rev.~A}\ }\textbf {\bibinfo {volume} {92}},\ \bibinfo
  {pages} {043410} (\bibinfo {year} {2015})}\BibitemShut {NoStop}%
\bibitem [{\citenamefont {Hohenester}\ \emph {et~al.}(2007)\citenamefont
  {Hohenester}, \citenamefont {Rekdal}, \citenamefont {{Borz\`{\i}}},\ and\
  \citenamefont {Schmiedmayer}}]{2007hohenester0a}%
  \BibitemOpen
  \bibfield  {author} {\bibinfo {author} {\bibfnamefont {U.}~\bibnamefont
  {Hohenester}}, \bibinfo {author} {\bibfnamefont {P.}~\bibnamefont {Rekdal}},
  \bibinfo {author} {\bibfnamefont {A.}~\bibnamefont {{Borz\`{\i}}}},\ and\
  \bibinfo {author} {\bibfnamefont {J.}~\bibnamefont {Schmiedmayer}},\
  }\bibfield  {title} {\bibinfo {title} {{Optimal quantum control of
  Bose-Einstein condensates in magnetic microtraps}},\ }\href
  {https://doi.org/10.1103/PhysRevA.75.023602} {\bibfield  {journal} {\bibinfo
  {journal} {Phys.~Rev.~A}\ }\textbf {\bibinfo {volume} {75}},\ \bibinfo
  {pages} {023602} (\bibinfo {year} {2007})}\BibitemShut {NoStop}%
\bibitem [{\citenamefont {{Gu{\'e}ry-Odelin}}\ and\ \citenamefont
  {Muga}(2014)}]{2014gueryodelin0a}%
  \BibitemOpen
  \bibfield  {author} {\bibinfo {author} {\bibfnamefont {D.}~\bibnamefont
  {{Gu{\'e}ry-Odelin}}}\ and\ \bibinfo {author} {\bibfnamefont {J.~G.}\
  \bibnamefont {Muga}},\ }\bibfield  {title} {\bibinfo {title} {{Transport in a
  harmonic trap: Shortcuts to adiabaticity and robust protocols}},\ }\href
  {https://doi.org/10.1103/PhysRevA.90.063425} {\bibfield  {journal} {\bibinfo
  {journal} {Phys.~Rev.~A.}\ }\textbf {\bibinfo {volume} {90}},\ \bibinfo
  {pages} {063425} (\bibinfo {year} {2014})}\BibitemShut {NoStop}%
\bibitem [{\citenamefont {Deffner}\ \emph {et~al.}(2014)\citenamefont
  {Deffner}, \citenamefont {Jarzynski},\ and\ \citenamefont {{del
  Campo}}}]{2014deffner0a}%
  \BibitemOpen
  \bibfield  {author} {\bibinfo {author} {\bibfnamefont {S.}~\bibnamefont
  {Deffner}}, \bibinfo {author} {\bibfnamefont {C.}~\bibnamefont {Jarzynski}},\
  and\ \bibinfo {author} {\bibfnamefont {A.}~\bibnamefont {{del Campo}}},\
  }\bibfield  {title} {\bibinfo {title} {{Classical and Quantum Shortcuts to
  Adiabaticity for Scale-Invariant Driving}},\ }\href
  {https://doi.org/10.1103/PhysRevX.4.021013} {\bibfield  {journal} {\bibinfo
  {journal} {Phys.~Rev.~X}\ }\textbf {\bibinfo {volume} {4}},\ \bibinfo {pages}
  {021013} (\bibinfo {year} {2014})}\BibitemShut {NoStop}%
\bibitem [{\citenamefont {Zhang}\ \emph {et~al.}(2016)\citenamefont {Zhang},
  \citenamefont {Muga}, \citenamefont {{Gu{\'e}ry-Odelin}},\ and\ \citenamefont
  {Chen}}]{2016zhang0a}%
  \BibitemOpen
  \bibfield  {author} {\bibinfo {author} {\bibfnamefont {Q.}~\bibnamefont
  {Zhang}}, \bibinfo {author} {\bibfnamefont {J.~G.}\ \bibnamefont {Muga}},
  \bibinfo {author} {\bibfnamefont {D.}~\bibnamefont {{Gu{\'e}ry-Odelin}}},\
  and\ \bibinfo {author} {\bibfnamefont {X.}~\bibnamefont {Chen}},\ }\bibfield
  {title} {\bibinfo {title} {{Optimal shortcuts for atomic transport in
  anharmonic traps}},\ }\href {https://doi.org/10.1088/0953-4075/49/12/125503}
  {\bibfield  {journal} {\bibinfo  {journal} {J.~Phys.~B}\ }\textbf {\bibinfo
  {volume} {49}},\ \bibinfo {pages} {125503} (\bibinfo {year}
  {2016})}\BibitemShut {NoStop}%
\bibitem [{\citenamefont {Amri}\ \emph {et~al.}(2019)\citenamefont {Amri},
  \citenamefont {Corgier}, \citenamefont {Sugny}, \citenamefont {Rasel},
  \citenamefont {Gaaloul},\ and\ \citenamefont {Charron}}]{2019amri0a}%
  \BibitemOpen
  \bibfield  {author} {\bibinfo {author} {\bibfnamefont {S.}~\bibnamefont
  {Amri}}, \bibinfo {author} {\bibfnamefont {R.}~\bibnamefont {Corgier}},
  \bibinfo {author} {\bibfnamefont {D.}~\bibnamefont {Sugny}}, \bibinfo
  {author} {\bibfnamefont {E.~M.}\ \bibnamefont {Rasel}}, \bibinfo {author}
  {\bibfnamefont {N.}~\bibnamefont {Gaaloul}},\ and\ \bibinfo {author}
  {\bibfnamefont {E.}~\bibnamefont {Charron}},\ }\bibfield  {title} {\bibinfo
  {title} {{Optimal control of the transport of Bose-Einstein condensates with
  atom chips}},\ }\href {https://doi.org/10.1038/s41598-019-41784-z} {\bibfield
   {journal} {\bibinfo  {journal} {Sci.~Rep.}\ }\textbf {\bibinfo {volume}
  {9}},\ \bibinfo {pages} {5346} (\bibinfo {year} {2019})}\BibitemShut
  {NoStop}%
\bibitem [{\citenamefont {Greiner}\ \emph {et~al.}(2001)\citenamefont
  {Greiner}, \citenamefont {Bloch}, \citenamefont {H\"ansch},\ and\
  \citenamefont {Esslinger}}]{2001greiner0a}%
  \BibitemOpen
  \bibfield  {author} {\bibinfo {author} {\bibfnamefont {M.}~\bibnamefont
  {Greiner}}, \bibinfo {author} {\bibfnamefont {I.}~\bibnamefont {Bloch}},
  \bibinfo {author} {\bibfnamefont {T.~W.}\ \bibnamefont {H\"ansch}},\ and\
  \bibinfo {author} {\bibfnamefont {T.}~\bibnamefont {Esslinger}},\ }\bibfield
  {title} {\bibinfo {title} {Magnetic transport of trapped cold atoms over a
  large distance},\ }\href {https://doi.org/10.1103/PhysRevA.63.031401}
  {\bibfield  {journal} {\bibinfo  {journal} {Phys. Rev. A}\ }\textbf {\bibinfo
  {volume} {63}},\ \bibinfo {pages} {031401} (\bibinfo {year}
  {2001})}\BibitemShut {NoStop}%
\bibitem [{\citenamefont {H{\"a}nsel}\ \emph {et~al.}(2001)\citenamefont
  {H{\"a}nsel}, \citenamefont {Reichel}, \citenamefont {Hommelhoff},\ and\
  \citenamefont {H{\"a}nsch}}]{2001hansel0b}%
  \BibitemOpen
  \bibfield  {author} {\bibinfo {author} {\bibfnamefont {W.}~\bibnamefont
  {H{\"a}nsel}}, \bibinfo {author} {\bibfnamefont {J.}~\bibnamefont {Reichel}},
  \bibinfo {author} {\bibfnamefont {P.}~\bibnamefont {Hommelhoff}},\ and\
  \bibinfo {author} {\bibfnamefont {T.~W.}\ \bibnamefont {H{\"a}nsch}},\
  }\bibfield  {title} {\bibinfo {title} {Magnetic conveyor belt for
  transporting and merging trapped atom clouds},\ }\href
  {https://doi.org/10.1103/PhysRevLett.86.608} {\bibfield  {journal} {\bibinfo
  {journal} {Phys.~Rev.~Lett.}\ }\textbf {\bibinfo {volume} {86}},\ \bibinfo
  {pages} {608} (\bibinfo {year} {2001})}\BibitemShut {NoStop}%
\bibitem [{\citenamefont {Minniberger}\ \emph {et~al.}(2014)\citenamefont
  {Minniberger}, \citenamefont {Diorice}, \citenamefont {Haslinger},
  \citenamefont {Hufnagel}, \citenamefont {Novotny}, \citenamefont {Lippok},
  \citenamefont {Majer}, \citenamefont {Koller}, \citenamefont {Schneider},\
  and\ \citenamefont {Schmiedmayer}}]{2014minniberger0a}%
  \BibitemOpen
  \bibfield  {author} {\bibinfo {author} {\bibfnamefont {S.}~\bibnamefont
  {Minniberger}}, \bibinfo {author} {\bibfnamefont {F.}~\bibnamefont
  {Diorice}}, \bibinfo {author} {\bibfnamefont {S.}~\bibnamefont {Haslinger}},
  \bibinfo {author} {\bibfnamefont {C.}~\bibnamefont {Hufnagel}}, \bibinfo
  {author} {\bibfnamefont {C.}~\bibnamefont {Novotny}}, \bibinfo {author}
  {\bibfnamefont {N.}~\bibnamefont {Lippok}}, \bibinfo {author} {\bibfnamefont
  {J.}~\bibnamefont {Majer}}, \bibinfo {author} {\bibfnamefont
  {C.}~\bibnamefont {Koller}}, \bibinfo {author} {\bibfnamefont
  {S.}~\bibnamefont {Schneider}},\ and\ \bibinfo {author} {\bibfnamefont
  {J.}~\bibnamefont {Schmiedmayer}},\ }\bibfield  {title} {\bibinfo {title}
  {Magnetic conveyor belt transport of ultracold atoms to a superconducting
  atomchip},\ }\href {https://doi.org/doi.org/10.1007/s00340-014-5790-5}
  {\bibfield  {journal} {\bibinfo  {journal} {Applied Physics B}\ }\textbf
  {\bibinfo {volume} {116}},\ \bibinfo {pages} {1017} (\bibinfo {year}
  {2014})}\BibitemShut {NoStop}%
\bibitem [{\citenamefont {Badr}\ \emph {et~al.}(2019)\citenamefont {Badr},
  \citenamefont {Ali}, \citenamefont {Seaward}, \citenamefont {Guo},
  \citenamefont {Wiotte}, \citenamefont {Dubessy}, \citenamefont {Perrin},\
  and\ \citenamefont {Perrin}}]{2019badr0a}%
  \BibitemOpen
  \bibfield  {author} {\bibinfo {author} {\bibfnamefont {T.}~\bibnamefont
  {Badr}}, \bibinfo {author} {\bibfnamefont {D.~B.}\ \bibnamefont {Ali}},
  \bibinfo {author} {\bibfnamefont {J.}~\bibnamefont {Seaward}}, \bibinfo
  {author} {\bibfnamefont {Y.}~\bibnamefont {Guo}}, \bibinfo {author}
  {\bibfnamefont {F.}~\bibnamefont {Wiotte}}, \bibinfo {author} {\bibfnamefont
  {R.}~\bibnamefont {Dubessy}}, \bibinfo {author} {\bibfnamefont
  {H.}~\bibnamefont {Perrin}},\ and\ \bibinfo {author} {\bibfnamefont
  {A.}~\bibnamefont {Perrin}},\ }\bibfield  {title} {\bibinfo {title}
  {Comparison of time profiles for the magnetic transport of cold atoms},\
  }\href {https://doi.org/10.1007/s00340-019-7213-0} {\bibfield  {journal}
  {\bibinfo  {journal} {Appl.~Phys.~B}\ }\textbf {\bibinfo {volume} {125}},\
  \bibinfo {pages} {102} (\bibinfo {year} {2019})}\BibitemShut {NoStop}%
\bibitem [{\citenamefont {Han}\ \emph {et~al.}(2013)\citenamefont {Han},
  \citenamefont {Xu}, \citenamefont {Zhang},\ and\ \citenamefont
  {Wang}}]{2013han0a}%
  \BibitemOpen
  \bibfield  {author} {\bibinfo {author} {\bibfnamefont {J.}~\bibnamefont
  {Han}}, \bibinfo {author} {\bibfnamefont {X.}~\bibnamefont {Xu}}, \bibinfo
  {author} {\bibfnamefont {H.}~\bibnamefont {Zhang}},\ and\ \bibinfo {author}
  {\bibfnamefont {Y.}~\bibnamefont {Wang}},\ }\bibfield  {title} {\bibinfo
  {title} {Optimal transport of cold atoms by modulating the velocity of
  traps},\ }\href {https://doi.org/10.1088/1674-1056/22/2/023702} {\bibfield
  {journal} {\bibinfo  {journal} {Chin.~Phys.~B}\ }\textbf {\bibinfo {volume}
  {22}},\ \bibinfo {pages} {023702} (\bibinfo {year} {2013})}\BibitemShut
  {NoStop}%
\bibitem [{\citenamefont {Chen}\ \emph {et~al.}(2010)\citenamefont {Chen},
  \citenamefont {Zhang}, \citenamefont {Xu}, \citenamefont {Li},\ and\
  \citenamefont {Wang}}]{2010chen0a}%
  \BibitemOpen
  \bibfield  {author} {\bibinfo {author} {\bibfnamefont {D.}~\bibnamefont
  {Chen}}, \bibinfo {author} {\bibfnamefont {H.}~\bibnamefont {Zhang}},
  \bibinfo {author} {\bibfnamefont {X.}~\bibnamefont {Xu}}, \bibinfo {author}
  {\bibfnamefont {T.}~\bibnamefont {Li}},\ and\ \bibinfo {author}
  {\bibfnamefont {Y.}~\bibnamefont {Wang}},\ }\bibfield  {title} {\bibinfo
  {title} {Nonadiabatic transport of cold atoms in a magnetic quadrupole
  potential},\ }\href {https://doi.org/10.1063/1.3377919} {\bibfield  {journal}
  {\bibinfo  {journal} {Appl.~Phys.~Lett.}\ }\textbf {\bibinfo {volume} {96}},\
  \bibinfo {pages} {134103} (\bibinfo {year} {2010})}\BibitemShut {NoStop}%
\bibitem [{\citenamefont {Ness}\ \emph {et~al.}(2018)\citenamefont {Ness},
  \citenamefont {Shkedrov}, \citenamefont {Florshaim},\ and\ \citenamefont
  {Y}}]{2018ness0a}%
  \BibitemOpen
  \bibfield  {author} {\bibinfo {author} {\bibfnamefont {G.}~\bibnamefont
  {Ness}}, \bibinfo {author} {\bibfnamefont {C.}~\bibnamefont {Shkedrov}},
  \bibinfo {author} {\bibfnamefont {Y.}~\bibnamefont {Florshaim}},\ and\
  \bibinfo {author} {\bibfnamefont {S.}~\bibnamefont {Y}},\ }\bibfield  {title}
  {\bibinfo {title} {{Realistic shortcuts to adiabaticity in optical
  transfer}},\ }\href {https://doi.org/10.1088/1367-2630/aadcc1} {\bibfield
  {journal} {\bibinfo  {journal} {New~J.~Phys.}\ }\textbf {\bibinfo {volume}
  {20}},\ \bibinfo {pages} {095002} (\bibinfo {year} {2018})}\BibitemShut
  {NoStop}%
\bibitem [{\citenamefont {Couvert}\ \emph {et~al.}(2008)\citenamefont
  {Couvert}, \citenamefont {Kawalec}, \citenamefont {Reinaudi},\ and\
  \citenamefont {Gu{\'{e}}ry-Odelin}}]{couvert08a}%
  \BibitemOpen
  \bibfield  {author} {\bibinfo {author} {\bibfnamefont {A.}~\bibnamefont
  {Couvert}}, \bibinfo {author} {\bibfnamefont {T.}~\bibnamefont {Kawalec}},
  \bibinfo {author} {\bibfnamefont {G.}~\bibnamefont {Reinaudi}},\ and\
  \bibinfo {author} {\bibfnamefont {D.}~\bibnamefont {Gu{\'{e}}ry-Odelin}},\
  }\bibfield  {title} {\bibinfo {title} {Optimal transport of ultracold atoms
  in the non-adiabatic regime},\ }\href
  {https://doi.org/10.1209/0295-5075/83/13001} {\bibfield  {journal} {\bibinfo
  {journal} {{EPL} (Europhysics Letters)}\ }\textbf {\bibinfo {volume} {83}},\
  \bibinfo {pages} {13001} (\bibinfo {year} {2008})}\BibitemShut {NoStop}%
\bibitem [{\citenamefont {Gustavson}\ \emph {et~al.}(2001)\citenamefont
  {Gustavson}, \citenamefont {Chikkatur}, \citenamefont {Leanhardt},
  \citenamefont {G\"orlitz}, \citenamefont {Gupta}, \citenamefont {Pritchard},\
  and\ \citenamefont {Ketterle}}]{2001gustavson0a}%
  \BibitemOpen
  \bibfield  {author} {\bibinfo {author} {\bibfnamefont {T.~L.}\ \bibnamefont
  {Gustavson}}, \bibinfo {author} {\bibfnamefont {A.~P.}\ \bibnamefont
  {Chikkatur}}, \bibinfo {author} {\bibfnamefont {A.~E.}\ \bibnamefont
  {Leanhardt}}, \bibinfo {author} {\bibfnamefont {A.}~\bibnamefont
  {G\"orlitz}}, \bibinfo {author} {\bibfnamefont {S.}~\bibnamefont {Gupta}},
  \bibinfo {author} {\bibfnamefont {D.~E.}\ \bibnamefont {Pritchard}},\ and\
  \bibinfo {author} {\bibfnamefont {W.}~\bibnamefont {Ketterle}},\ }\bibfield
  {title} {\bibinfo {title} {Transport of bose-einstein condensates with
  optical tweezers},\ }\href {https://doi.org/10.1103/PhysRevLett.88.020401}
  {\bibfield  {journal} {\bibinfo  {journal} {Phys. Rev. Lett.}\ }\textbf
  {\bibinfo {volume} {88}},\ \bibinfo {pages} {020401} (\bibinfo {year}
  {2001})}\BibitemShut {NoStop}%
\bibitem [{\citenamefont {L{\'{e}}onard}\ \emph {et~al.}(2014)\citenamefont
  {L{\'{e}}onard}, \citenamefont {Lee}, \citenamefont {Morales}, \citenamefont
  {Karg}, \citenamefont {Esslinger},\ and\ \citenamefont
  {Donner}}]{leonard14a}%
  \BibitemOpen
  \bibfield  {author} {\bibinfo {author} {\bibfnamefont {J.}~\bibnamefont
  {L{\'{e}}onard}}, \bibinfo {author} {\bibfnamefont {M.}~\bibnamefont {Lee}},
  \bibinfo {author} {\bibfnamefont {A.}~\bibnamefont {Morales}}, \bibinfo
  {author} {\bibfnamefont {T.~M.}\ \bibnamefont {Karg}}, \bibinfo {author}
  {\bibfnamefont {T.}~\bibnamefont {Esslinger}},\ and\ \bibinfo {author}
  {\bibfnamefont {T.}~\bibnamefont {Donner}},\ }\bibfield  {title} {\bibinfo
  {title} {Optical transport and manipulation of an ultracold atomic cloud
  using focus-tunable lenses},\ }\href
  {https://doi.org/10.1088/1367-2630/16/9/093028} {\bibfield  {journal}
  {\bibinfo  {journal} {New Journal of Physics}\ }\textbf {\bibinfo {volume}
  {16}},\ \bibinfo {pages} {093028} (\bibinfo {year} {2014})}\BibitemShut
  {NoStop}%
\bibitem [{\citenamefont {Middelmann}\ \emph {et~al.}(2012)\citenamefont
  {Middelmann}, \citenamefont {Falke}, \citenamefont {Lisdat},\ and\
  \citenamefont {Sterr}}]{2012middelmann0a}%
  \BibitemOpen
  \bibfield  {author} {\bibinfo {author} {\bibfnamefont {T.}~\bibnamefont
  {Middelmann}}, \bibinfo {author} {\bibfnamefont {S.}~\bibnamefont {Falke}},
  \bibinfo {author} {\bibfnamefont {C.}~\bibnamefont {Lisdat}},\ and\ \bibinfo
  {author} {\bibfnamefont {U.}~\bibnamefont {Sterr}},\ }\bibfield  {title}
  {\bibinfo {title} {Long-range transport of ultracold atoms in a far-detuned
  one-dimensional optical lattice},\ }\href
  {https://doi.org/10.1088/1367-2630/14/7/073020} {\bibfield  {journal}
  {\bibinfo  {journal} {New.~J.~Phys.}\ }\textbf {\bibinfo {volume} {14}},\
  \bibinfo {pages} {073020} (\bibinfo {year} {2012})}\BibitemShut {NoStop}%
\bibitem [{\citenamefont {Schmid}\ \emph {et~al.}(2006)\citenamefont {Schmid},
  \citenamefont {Thalhammer}, \citenamefont {Winkler}, \citenamefont {Lang},\
  and\ \citenamefont {{Hecker Denschlag}}}]{2006schmid0a}%
  \BibitemOpen
  \bibfield  {author} {\bibinfo {author} {\bibfnamefont {S.}~\bibnamefont
  {Schmid}}, \bibinfo {author} {\bibfnamefont {G.}~\bibnamefont {Thalhammer}},
  \bibinfo {author} {\bibfnamefont {K.}~\bibnamefont {Winkler}}, \bibinfo
  {author} {\bibfnamefont {F.}~\bibnamefont {Lang}},\ and\ \bibinfo {author}
  {\bibfnamefont {J.}~\bibnamefont {{Hecker Denschlag}}},\ }\bibfield  {title}
  {\bibinfo {title} {{Long distance transport of ultracold atoms using a 1D
  optical lattice}},\ }\href {https://doi.org/10.1088/1367-2630/8/8/159}
  {\bibfield  {journal} {\bibinfo  {journal} {New~J.~Phys.}\ }\textbf {\bibinfo
  {volume} {8}},\ \bibinfo {pages} {159} (\bibinfo {year} {2006})}\BibitemShut
  {NoStop}%
\bibitem [{\citenamefont {Mandel}\ \emph {et~al.}(2003)\citenamefont {Mandel},
  \citenamefont {Greiner}, \citenamefont {Widera}, \citenamefont {Rom},
  \citenamefont {H{\"a}nsch},\ and\ \citenamefont {Bloch}}]{2003mandel0a}%
  \BibitemOpen
  \bibfield  {author} {\bibinfo {author} {\bibfnamefont {O.}~\bibnamefont
  {Mandel}}, \bibinfo {author} {\bibfnamefont {M.}~\bibnamefont {Greiner}},
  \bibinfo {author} {\bibfnamefont {A.}~\bibnamefont {Widera}}, \bibinfo
  {author} {\bibfnamefont {T.}~\bibnamefont {Rom}}, \bibinfo {author}
  {\bibfnamefont {T.~W.}\ \bibnamefont {H{\"a}nsch}},\ and\ \bibinfo {author}
  {\bibfnamefont {I.}~\bibnamefont {Bloch}},\ }\bibfield  {title} {\bibinfo
  {title} {Coherent transport of neutral atoms in spin-dependent optical
  lattice potentials},\ }\href {https://doi.org/10.1103/PhysRevLett.91.010407}
  {\bibfield  {journal} {\bibinfo  {journal} {Phys.~Rev.~Lett.}\ }\textbf
  {\bibinfo {volume} {91}},\ \bibinfo {pages} {010407} (\bibinfo {year}
  {2003})}\BibitemShut {NoStop}%
\bibitem [{\citenamefont {Browaeys}\ \emph {et~al.}(2005)\citenamefont
  {Browaeys}, \citenamefont {H{\"a}ffner}, \citenamefont {{McKenzie}},
  \citenamefont {Rolston}, \citenamefont {Helmerson},\ and\ \citenamefont
  {Phillips}}]{2005browaeys0a}%
  \BibitemOpen
  \bibfield  {author} {\bibinfo {author} {\bibfnamefont {A.}~\bibnamefont
  {Browaeys}}, \bibinfo {author} {\bibfnamefont {H.}~\bibnamefont
  {H{\"a}ffner}}, \bibinfo {author} {\bibfnamefont {C.}~\bibnamefont
  {{McKenzie}}}, \bibinfo {author} {\bibfnamefont {S.~L.}\ \bibnamefont
  {Rolston}}, \bibinfo {author} {\bibfnamefont {K.}~\bibnamefont {Helmerson}},\
  and\ \bibinfo {author} {\bibfnamefont {W.~D.}\ \bibnamefont {Phillips}},\
  }\bibfield  {title} {\bibinfo {title} {{Transport of atoms in a quantum
  conveyor belt}},\ }\href {https://doi.org/10.1103/PhysRevA.72.053605}
  {\bibfield  {journal} {\bibinfo  {journal} {Phys.~Rev.~A}\ }\textbf {\bibinfo
  {volume} {72}},\ \bibinfo {pages} {053605} (\bibinfo {year}
  {2005})}\BibitemShut {NoStop}%
\bibitem [{\citenamefont {Corgier}\ \emph {et~al.}(2018)\citenamefont
  {Corgier}, \citenamefont {Amri}, \citenamefont {Herr}, \citenamefont
  {Ahlers}, \citenamefont {Rudolph}, \citenamefont {{Gu{\'e}ry-Odelin}},
  \citenamefont {Rasel}, \citenamefont {Charron},\ and\ \citenamefont
  {Gaaloul}}]{2018corgier0a}%
  \BibitemOpen
  \bibfield  {author} {\bibinfo {author} {\bibfnamefont {R.}~\bibnamefont
  {Corgier}}, \bibinfo {author} {\bibfnamefont {S.}~\bibnamefont {Amri}},
  \bibinfo {author} {\bibfnamefont {W.}~\bibnamefont {Herr}}, \bibinfo {author}
  {\bibfnamefont {H.}~\bibnamefont {Ahlers}}, \bibinfo {author} {\bibfnamefont
  {J.}~\bibnamefont {Rudolph}}, \bibinfo {author} {\bibfnamefont
  {D.}~\bibnamefont {{Gu{\'e}ry-Odelin}}}, \bibinfo {author} {\bibfnamefont
  {E.~M.}\ \bibnamefont {Rasel}}, \bibinfo {author} {\bibfnamefont
  {E.}~\bibnamefont {Charron}},\ and\ \bibinfo {author} {\bibfnamefont
  {N.}~\bibnamefont {Gaaloul}},\ }\bibfield  {title} {\bibinfo {title} {{Fast
  manipulation of Bose-Einstein condensates with an atom chip}},\ }\href
  {https://doi.org/10.1088/1367-2630/aabdfc} {\bibfield  {journal} {\bibinfo
  {journal} {New~J.~Phys.}\ }\textbf {\bibinfo {volume} {20}},\ \bibinfo
  {pages} {055002} (\bibinfo {year} {2018})}\BibitemShut {NoStop}%
\bibitem [{\citenamefont {Husimi}(1953)}]{1953husimi0a}%
  \BibitemOpen
  \bibfield  {author} {\bibinfo {author} {\bibfnamefont {K.}~\bibnamefont
  {Husimi}},\ }\bibfield  {title} {\bibinfo {title} {{Miscellanea in Elementary
  Quantum Mechanics, II}},\ }\href {https://doi.org/10.1143/ptp/9.4.381}
  {\bibfield  {journal} {\bibinfo  {journal} {Progress of Theoretical Physics}\
  }\textbf {\bibinfo {volume} {9}},\ \bibinfo {pages} {381} (\bibinfo {year}
  {1953})},\ \bibinfo {note} {with contributions noted from T.~Taniuti,
  Y.~Suzuki, T.~Dodo, M.~{\^O}tuka, R.~Utiyama, and H.~Arita}\BibitemShut
  {NoStop}%
\bibitem [{\citenamefont {Kleber}(1994)}]{1994kleber0a}%
  \BibitemOpen
  \bibfield  {author} {\bibinfo {author} {\bibfnamefont {M.}~\bibnamefont
  {Kleber}},\ }\bibfield  {title} {\bibinfo {title} {{Exact solutions for
  time-dependent phenomena in quantum mechanics}},\ }\href
  {https://doi.org/10.1016/0370-1573(94)90029-9} {\bibfield  {journal}
  {\bibinfo  {journal} {Phys.~Rep.}\ }\textbf {\bibinfo {volume} {236}},\
  \bibinfo {pages} {331} (\bibinfo {year} {1994})}\BibitemShut {NoStop}%
\bibitem [{\citenamefont {Kagan}\ \emph {et~al.}(1996)\citenamefont {Kagan},
  \citenamefont {Surkov},\ and\ \citenamefont {Shlyapnikov}}]{1996kagan0a}%
  \BibitemOpen
  \bibfield  {author} {\bibinfo {author} {\bibfnamefont {Y.}~\bibnamefont
  {Kagan}}, \bibinfo {author} {\bibfnamefont {E.~L.}\ \bibnamefont {Surkov}},\
  and\ \bibinfo {author} {\bibfnamefont {G.~V.}\ \bibnamefont {Shlyapnikov}},\
  }\bibfield  {title} {\bibinfo {title} {Evolution of a bose-condensed gas
  under variations of the confining potential},\ }\href
  {https://doi.org/10.1103/PhysRevA.54.R1753} {\bibfield  {journal} {\bibinfo
  {journal} {Phys.~Rev.~A}\ }\textbf {\bibinfo {volume} {54}},\ \bibinfo
  {pages} {R1753} (\bibinfo {year} {1996})}\BibitemShut {NoStop}%
\bibitem [{\citenamefont {Feynman}(1950)}]{1950feynman0a}%
  \BibitemOpen
  \bibfield  {author} {\bibinfo {author} {\bibfnamefont {R.~P.}\ \bibnamefont
  {Feynman}},\ }\bibfield  {title} {\bibinfo {title} {{Mathematical Formulation
  of the Quantum Theory of Electromagnetic Interaction}},\ }\href
  {https://doi.org/10.1103/PhysRev.80.440} {\bibfield  {journal} {\bibinfo
  {journal} {Phys.~Rev.}\ }\textbf {\bibinfo {volume} {80}},\ \bibinfo {pages}
  {440} (\bibinfo {year} {1950})}\BibitemShut {NoStop}%
\bibitem [{\citenamefont {Dykhne}(1960)}]{1960dykhne0a}%
  \BibitemOpen
  \bibfield  {author} {\bibinfo {author} {\bibfnamefont {A.~M.}\ \bibnamefont
  {Dykhne}},\ }\bibfield  {title} {\bibinfo {title} {{Quantum transitions in
  the adiabatic approximation}},\ }\href
  {http://jetp.ras.ru/cgi-bin/e/index/r/38/2/p570?a=list} {\bibfield  {journal}
  {\bibinfo  {journal} {Sov.~Phys.~JETP}\ }\textbf {\bibinfo {volume} {11}},\
  \bibinfo {pages} {411} (\bibinfo {year} {1960})},\ \bibinfo {note}
  {zh.~Eksp.~Teor.~Fiz. 38, 570 (1960)}\BibitemShut {NoStop}%
\bibitem [{\citenamefont {Solimeno}\ \emph {et~al.}(1969)\citenamefont
  {Solimeno}, \citenamefont {{Di Porto}},\ and\ \citenamefont
  {Crosignani}}]{1969solimeno0a}%
  \BibitemOpen
  \bibfield  {author} {\bibinfo {author} {\bibfnamefont {S.}~\bibnamefont
  {Solimeno}}, \bibinfo {author} {\bibfnamefont {P.}~\bibnamefont {{Di
  Porto}}},\ and\ \bibinfo {author} {\bibfnamefont {B.}~\bibnamefont
  {Crosignani}},\ }\bibfield  {title} {\bibinfo {title} {{Quantum Harmonic
  Oscillator with Time‐Dependent Frequency}},\ }\href
  {https://doi.org/10.1063/1.1664783} {\bibfield  {journal} {\bibinfo
  {journal} {J.~Math.~Phys.}\ }\textbf {\bibinfo {volume} {10}},\ \bibinfo
  {pages} {1922} (\bibinfo {year} {1969})}\BibitemShut {NoStop}%
\bibitem [{\citenamefont {Popov}\ and\ \citenamefont
  {Perelomov}(1969)}]{1969popov0a}%
  \BibitemOpen
  \bibfield  {author} {\bibinfo {author} {\bibfnamefont {V.~S.}\ \bibnamefont
  {Popov}}\ and\ \bibinfo {author} {\bibfnamefont {A.~M.}\ \bibnamefont
  {Perelomov}},\ }\bibfield  {title} {\bibinfo {title} {{Parametric Excitation
  of a Quantum Oscillator}},\ }\href
  {http://jetp.ras.ru/cgi-bin/e/index/e/29/4/p738?a=list} {\bibfield  {journal}
  {\bibinfo  {journal} {Sov.~Phys.~JETP}\ }\textbf {\bibinfo {volume} {29}},\
  \bibinfo {pages} {738} (\bibinfo {year} {1969})},\ \bibinfo {note}
  {zh.~Eksp.~Teor.~Fiz. 56, 1375 (1969)}\BibitemShut {NoStop}%
\bibitem [{\citenamefont {Lohe}(2009)}]{2009lohe0a}%
  \BibitemOpen
  \bibfield  {author} {\bibinfo {author} {\bibfnamefont {M.~A.}\ \bibnamefont
  {Lohe}},\ }\bibfield  {title} {\bibinfo {title} {{Exact time dependence of
  solutions to the time-dependent Schr{\"o}dinger equation}},\ }\href
  {https://doi.org/10.1088/1751-8113/42/3/035307} {\bibfield  {journal}
  {\bibinfo  {journal} {J.~Math.~Phys.~A}\ }\textbf {\bibinfo {volume} {42}},\
  \bibinfo {pages} {035307} (\bibinfo {year} {2009})}\BibitemShut {NoStop}%
\bibitem [{\citenamefont {Feynman}(1951)}]{1951feynman0a}%
  \BibitemOpen
  \bibfield  {author} {\bibinfo {author} {\bibfnamefont {R.~P.}\ \bibnamefont
  {Feynman}},\ }\bibfield  {title} {\bibinfo {title} {{An Operator Calculus
  Having Applications in Quantum Electrodynamics}},\ }\href
  {https://doi.org/10.1103/PhysRev.84.108} {\bibfield  {journal} {\bibinfo
  {journal} {Phys.~Rev.}\ }\textbf {\bibinfo {volume} {84}},\ \bibinfo {pages}
  {108} (\bibinfo {year} {1951})}\BibitemShut {NoStop}%
\bibitem [{\citenamefont {Fujiwara}(1952)}]{1952fujiwara0a}%
  \BibitemOpen
  \bibfield  {author} {\bibinfo {author} {\bibfnamefont {I.}~\bibnamefont
  {Fujiwara}},\ }\bibfield  {title} {\bibinfo {title} {{Operator Calculus of
  Quantized Operator}},\ }\href {https://doi.org/10.1143/PTP.7.5.433}
  {\bibfield  {journal} {\bibinfo  {journal} {Prog.~Theor.~Phys.}\ }\textbf
  {\bibinfo {volume} {7}},\ \bibinfo {pages} {433} (\bibinfo {year}
  {1952})}\BibitemShut {NoStop}%
\bibitem [{\citenamefont {Kerner}(1958)}]{1958kerner0a}%
  \BibitemOpen
  \bibfield  {author} {\bibinfo {author} {\bibfnamefont {E.~H.}\ \bibnamefont
  {Kerner}},\ }\bibfield  {title} {\bibinfo {title} {{Note on the Forced and
  Damped Oscillator in Quantum Mechanics}},\ }\href
  {https://doi.org/10.1139/p58-038} {\bibfield  {journal} {\bibinfo  {journal}
  {Can.~J.~Phys.}\ }\textbf {\bibinfo {volume} {36}},\ \bibinfo {pages} {371}
  (\bibinfo {year} {1958})}\BibitemShut {NoStop}%
\bibitem [{\citenamefont {Jackson}\ \emph {et~al.}(2021)\citenamefont
  {Jackson}, \citenamefont {Corney},\ and\ \citenamefont
  {Bromley}}]{2021jackson0a}%
  \BibitemOpen
  \bibfield  {author} {\bibinfo {author} {\bibfnamefont {C.~D.}\ \bibnamefont
  {Jackson}}, \bibinfo {author} {\bibfnamefont {J.~F.}\ \bibnamefont
  {Corney}},\ and\ \bibinfo {author} {\bibfnamefont {M.~W.~J.}\ \bibnamefont
  {Bromley}},\ }\href@noop {} {\bibinfo {title} {{Benchmarking numerical
  solutions to the doubly-time-dependent Schr\"{o}dinger equation}}} (\bibinfo
  {year} {2021}),\ \bibinfo {note} {in preparation}\BibitemShut {NoStop}%
\bibitem [{\citenamefont {Bandrauk}\ and\ \citenamefont
  {Shen}(1991)}]{1991bandrauk0a}%
  \BibitemOpen
  \bibfield  {author} {\bibinfo {author} {\bibfnamefont {A.~D.}\ \bibnamefont
  {Bandrauk}}\ and\ \bibinfo {author} {\bibfnamefont {H.}~\bibnamefont
  {Shen}},\ }\bibfield  {title} {\bibinfo {title} {{Improved exponential split
  operator method for solving the time-dependent Schr{\"o}dinger equation}},\
  }\href {https://doi.org/10.1016/0009-2614(91)90232-X} {\bibfield  {journal}
  {\bibinfo  {journal} {Chem.~Phys.~Lett.}\ }\textbf {\bibinfo {volume}
  {176}},\ \bibinfo {pages} {428} (\bibinfo {year} {1991})}\BibitemShut
  {NoStop}%
\bibitem [{\citenamefont {H{\"a}nggi}(1998)}]{1998hanggi0a}%
  \BibitemOpen
  \bibfield  {author} {\bibinfo {author} {\bibfnamefont {P.}~\bibnamefont
  {H{\"a}nggi}},\ }\bibinfo {title} {{Driven Quantum Systems}},\ Chap.~\bibinfo
  {chapter} {5},\ in\  \cite{1998dittrich0a} (\bibinfo {year}
  {1998})\BibitemShut {NoStop}%
\bibitem [{\citenamefont {Dittrich}\ \emph {et~al.}(1998)\citenamefont
  {Dittrich}, \citenamefont {H{\"a}nggi}, \citenamefont {Ingold}, \citenamefont
  {Kramer}, \citenamefont {Sch{\"o}n},\ and\ \citenamefont
  {Zwerger}}]{1998dittrich0a}%
  \BibitemOpen
  \bibinfo {editor} {\bibfnamefont {T.}~\bibnamefont {Dittrich}}, \bibinfo
  {editor} {\bibfnamefont {P.}~\bibnamefont {H{\"a}nggi}}, \bibinfo {editor}
  {\bibfnamefont {G.}~\bibnamefont {Ingold}}, \bibinfo {editor} {\bibfnamefont
  {B.}~\bibnamefont {Kramer}}, \bibinfo {editor} {\bibfnamefont
  {G.}~\bibnamefont {Sch{\"o}n}},\ and\ \bibinfo {editor} {\bibfnamefont
  {W.}~\bibnamefont {Zwerger}},\ eds.,\ \href@noop {} {\emph {\bibinfo {title}
  {{Quantum Transport and Dissipation}}}}\ (\bibinfo  {publisher} {Wiley-Vch},\
  \bibinfo {address} {Weinheim},\ \bibinfo {year} {1998})\BibitemShut {NoStop}%
\bibitem [{sup()}]{supplementary1}%
  \BibitemOpen
  \href@noop {} {}\bibinfo {note} {See Supplemental Material at [URL will be
  inserted by publisher] for technical details.}\BibitemShut {Stop}%
\bibitem [{\citenamefont {Dobson}(1994)}]{1994dobson0a}%
  \BibitemOpen
  \bibfield  {author} {\bibinfo {author} {\bibfnamefont {J.~F.}\ \bibnamefont
  {Dobson}},\ }\bibfield  {title} {\bibinfo {title} {{Harmonic-Potential
  Theorem: Implications for Approximate Many-Body Theories}},\ }\href
  {https://doi.org/10.1103/PhysRevLett.73.2244} {\bibfield  {journal} {\bibinfo
   {journal} {Phys.~Rev.~Lett.}\ }\textbf {\bibinfo {volume} {73}},\ \bibinfo
  {pages} {2244} (\bibinfo {year} {1994})}\BibitemShut {NoStop}%
\bibitem [{\citenamefont {Esry}(1997)}]{1997esry0a}%
  \BibitemOpen
  \bibfield  {author} {\bibinfo {author} {\bibfnamefont {B.~D.}\ \bibnamefont
  {Esry}},\ }\bibfield  {title} {\bibinfo {title} {{Hartree-Fock theory for
  Bose-Einstein condensates and the inclusion of correlation effects}},\ }\href
  {https://doi.org/10.1103/PhysRevA.55.1147} {\bibfield  {journal} {\bibinfo
  {journal} {Phys.~Rev.~A}\ }\textbf {\bibinfo {volume} {55}},\ \bibinfo
  {pages} {1147} (\bibinfo {year} {1997})}\BibitemShut {NoStop}%
\bibitem [{\citenamefont {Bell}\ \emph {et~al.}(2016)\citenamefont {Bell},
  \citenamefont {Glidden}, \citenamefont {Humbert}, \citenamefont {Bromley},
  \citenamefont {Haine}, \citenamefont {Davis}, \citenamefont {Neely},
  \citenamefont {Baker},\ and\ \citenamefont
  {{Rubinsztein-Dunlop}}}]{2016bell0a}%
  \BibitemOpen
  \bibfield  {author} {\bibinfo {author} {\bibfnamefont {T.~A.}\ \bibnamefont
  {Bell}}, \bibinfo {author} {\bibfnamefont {J.~A.~P.}\ \bibnamefont
  {Glidden}}, \bibinfo {author} {\bibfnamefont {L.}~\bibnamefont {Humbert}},
  \bibinfo {author} {\bibfnamefont {M.~W.~J.}\ \bibnamefont {Bromley}},
  \bibinfo {author} {\bibfnamefont {S.~A.}\ \bibnamefont {Haine}}, \bibinfo
  {author} {\bibfnamefont {M.~J.}\ \bibnamefont {Davis}}, \bibinfo {author}
  {\bibfnamefont {T.~W.}\ \bibnamefont {Neely}}, \bibinfo {author}
  {\bibfnamefont {M.~A.}\ \bibnamefont {Baker}},\ and\ \bibinfo {author}
  {\bibfnamefont {H.}~\bibnamefont {{Rubinsztein-Dunlop}}},\ }\bibfield
  {title} {\bibinfo {title} {{Bose-Einstein condensation in large time-averaged
  optical ring potentials}},\ }\href
  {https://doi.org/10.1088/1367-2630/18/3/035003} {\bibfield  {journal}
  {\bibinfo  {journal} {New~J.~Phys.}\ }\textbf {\bibinfo {volume} {18}},\
  \bibinfo {pages} {035003} (\bibinfo {year} {2016})}\BibitemShut {NoStop}%
\bibitem [{\citenamefont {Bell}\ \emph {et~al.}(2018)\citenamefont {Bell},
  \citenamefont {Gauthier}, \citenamefont {Neely}, \citenamefont
  {{Rubinsztein-Dunlop}}, \citenamefont {Davis},\ and\ \citenamefont
  {Baker}}]{2018bell0a}%
  \BibitemOpen
  \bibfield  {author} {\bibinfo {author} {\bibfnamefont {T.~A.}\ \bibnamefont
  {Bell}}, \bibinfo {author} {\bibfnamefont {G.}~\bibnamefont {Gauthier}},
  \bibinfo {author} {\bibfnamefont {T.~W.}\ \bibnamefont {Neely}}, \bibinfo
  {author} {\bibfnamefont {H.}~\bibnamefont {{Rubinsztein-Dunlop}}}, \bibinfo
  {author} {\bibfnamefont {M.~J.}\ \bibnamefont {Davis}},\ and\ \bibinfo
  {author} {\bibfnamefont {M.~A.}\ \bibnamefont {Baker}},\ }\bibfield  {title}
  {\bibinfo {title} {{Phase and micromotion of Bose-Einstein condensates in a
  time-averaged ring trap}},\ }\href
  {https://doi.org/10.1103/PhysRevA.98.013604} {\bibfield  {journal} {\bibinfo
  {journal} {Phys.~Rev.~A}\ }\textbf {\bibinfo {volume} {98}},\ \bibinfo
  {pages} {013604} (\bibinfo {year} {2018})}\BibitemShut {NoStop}%
\bibitem [{\citenamefont {Stickney}\ \emph {et~al.}(2017)\citenamefont
  {Stickney}, \citenamefont {Imhof}, \citenamefont {Kasch}, \citenamefont
  {Kroese}, \citenamefont {Crow}, \citenamefont {Olson},\ and\ \citenamefont
  {Squires}}]{2017stickney0a}%
  \BibitemOpen
  \bibfield  {author} {\bibinfo {author} {\bibfnamefont {J.~A.}\ \bibnamefont
  {Stickney}}, \bibinfo {author} {\bibfnamefont {E.}~\bibnamefont {Imhof}},
  \bibinfo {author} {\bibfnamefont {B.}~\bibnamefont {Kasch}}, \bibinfo
  {author} {\bibfnamefont {B.}~\bibnamefont {Kroese}}, \bibinfo {author}
  {\bibfnamefont {J.~A.~R.}\ \bibnamefont {Crow}}, \bibinfo {author}
  {\bibfnamefont {S.~E.}\ \bibnamefont {Olson}},\ and\ \bibinfo {author}
  {\bibfnamefont {M.~B.}\ \bibnamefont {Squires}},\ }\bibfield  {title}
  {\bibinfo {title} {{Tunable axial potentials for atom-chip waveguides}},\
  }\href {https://doi.org/10.1103/PhysRevA.96.053606} {\bibfield  {journal}
  {\bibinfo  {journal} {Phys.~Rev.~A}\ }\textbf {\bibinfo {volume} {96}},\
  \bibinfo {pages} {053606} (\bibinfo {year} {2017})}\BibitemShut {NoStop}%
\bibitem [{\citenamefont {Squires}\ \emph {et~al.}(2016)\citenamefont
  {Squires}, \citenamefont {Olson}, \citenamefont {Kasch}, \citenamefont
  {Stickney}, \citenamefont {Erickson}, \citenamefont {Crow}, \citenamefont
  {Carlson},\ and\ \citenamefont {Burke}}]{2016squires0a}%
  \BibitemOpen
  \bibfield  {author} {\bibinfo {author} {\bibfnamefont {M.~B.}\ \bibnamefont
  {Squires}}, \bibinfo {author} {\bibfnamefont {S.~E.}\ \bibnamefont {Olson}},
  \bibinfo {author} {\bibfnamefont {B.}~\bibnamefont {Kasch}}, \bibinfo
  {author} {\bibfnamefont {J.~A.}\ \bibnamefont {Stickney}}, \bibinfo {author}
  {\bibfnamefont {C.~J.}\ \bibnamefont {Erickson}}, \bibinfo {author}
  {\bibfnamefont {J.~A.~R.}\ \bibnamefont {Crow}}, \bibinfo {author}
  {\bibfnamefont {E.~J.}\ \bibnamefont {Carlson}},\ and\ \bibinfo {author}
  {\bibfnamefont {J.~H.}\ \bibnamefont {Burke}},\ }\bibfield  {title} {\bibinfo
  {title} {{Ex vacuo atom chip Bose-Einstein condensate}},\ }\href
  {https://doi.org/10.1063/1.4971838} {\bibfield  {journal} {\bibinfo
  {journal} {Appl.~Phys.~Lett.}\ }\textbf {\bibinfo {volume} {109}},\ \bibinfo
  {pages} {264101} (\bibinfo {year} {2016})}\BibitemShut {NoStop}%
\bibitem [{\citenamefont {Moan}\ \emph {et~al.}(2020)\citenamefont {Moan},
  \citenamefont {Horne}, \citenamefont {Arpornthip}, \citenamefont {Luo},
  \citenamefont {Fallon}, \citenamefont {Berl},\ and\ \citenamefont
  {Sackett}}]{2020moan0a}%
  \BibitemOpen
  \bibfield  {author} {\bibinfo {author} {\bibfnamefont {E.~R.}\ \bibnamefont
  {Moan}}, \bibinfo {author} {\bibfnamefont {R.~A.}\ \bibnamefont {Horne}},
  \bibinfo {author} {\bibfnamefont {T.}~\bibnamefont {Arpornthip}}, \bibinfo
  {author} {\bibfnamefont {Z.}~\bibnamefont {Luo}}, \bibinfo {author}
  {\bibfnamefont {A.~J.}\ \bibnamefont {Fallon}}, \bibinfo {author}
  {\bibfnamefont {S.~J.}\ \bibnamefont {Berl}},\ and\ \bibinfo {author}
  {\bibfnamefont {C.~A.}\ \bibnamefont {Sackett}},\ }\bibfield  {title}
  {\bibinfo {title} {{Quantum Rotation Sensing with Dual Sagnac Interferometers
  in an Atom-Optical Waveguide}},\ }\href
  {https://doi.org/10.1103/PhysRevLett.124.120403} {\bibfield  {journal}
  {\bibinfo  {journal} {Phys.~Rev.~Lett.}\ }\textbf {\bibinfo {volume} {124}},\
  \bibinfo {pages} {120403} (\bibinfo {year} {2020})}\BibitemShut {NoStop}%
\end{thebibliography}%


\begin{thebibliography}{7}%
\makeatletter
\providecommand \@ifxundefined [1]{%
 \@ifx{#1\undefined}
}%
\providecommand \@ifnum [1]{%
 \ifnum #1\expandafter \@firstoftwo
 \else \expandafter \@secondoftwo
 \fi
}%
\providecommand \@ifx [1]{%
 \ifx #1\expandafter \@firstoftwo
 \else \expandafter \@secondoftwo
 \fi
}%
\providecommand \natexlab [1]{#1}%
\providecommand \enquote  [1]{``#1''}%
\providecommand \bibnamefont  [1]{#1}%
\providecommand \bibfnamefont [1]{#1}%
\providecommand \citenamefont [1]{#1}%
\providecommand \href@noop [0]{\@secondoftwo}%
\providecommand \href [0]{\begingroup \@sanitize@url \@href}%
\providecommand \@href[1]{\@@startlink{#1}\@@href}%
\providecommand \@@href[1]{\endgroup#1\@@endlink}%
\providecommand \@sanitize@url [0]{\catcode `\\12\catcode `\$12\catcode
  `\&12\catcode `\#12\catcode `\^12\catcode `\_12\catcode `\%12\relax}%
\providecommand \@@startlink[1]{}%
\providecommand \@@endlink[0]{}%
\providecommand \url  [0]{\begingroup\@sanitize@url \@url }%
\providecommand \@url [1]{\endgroup\@href {#1}{\urlprefix }}%
\providecommand \urlprefix  [0]{URL }%
\providecommand \Eprint [0]{\href }%
\providecommand \doibase [0]{https://doi.org/}%
\providecommand \selectlanguage [0]{\@gobble}%
\providecommand \bibinfo  [0]{\@secondoftwo}%
\providecommand \bibfield  [0]{\@secondoftwo}%
\providecommand \translation [1]{[#1]}%
\providecommand \BibitemOpen [0]{}%
\providecommand \bibitemStop [0]{}%
\providecommand \bibitemNoStop [0]{.\EOS\space}%
\providecommand \EOS [0]{\spacefactor3000\relax}%
\providecommand \BibitemShut  [1]{\csname bibitem#1\endcsname}%
\let\auto@bib@innerbib\@empty
%</preamble>
\bibitem [{\citenamefont {Husimi}(1953)}]{1953husimi0a}%
  \BibitemOpen
  \bibfield  {author} {\bibinfo {author} {\bibfnamefont {K.}~\bibnamefont
  {Husimi}},\ }\bibfield  {title} {\bibinfo {title} {{Miscellanea in Elementary
  Quantum Mechanics, II}},\ }\href {https://doi.org/10.1143/ptp/9.4.381}
  {\bibfield  {journal} {\bibinfo  {journal} {Progress of Theoretical Physics}\
  }\textbf {\bibinfo {volume} {9}},\ \bibinfo {pages} {381} (\bibinfo {year}
  {1953})},\ \bibinfo {note} {with contributions noted from T.~Taniuti,
  Y.~Suzuki, T.~Dodo, M.~{\^O}tuka, R.~Utiyama, and H.~Arita}\BibitemShut
  {NoStop}%
\bibitem [{\citenamefont {Esry}(1997)}]{1997esry0a}%
  \BibitemOpen
  \bibfield  {author} {\bibinfo {author} {\bibfnamefont {B.~D.}\ \bibnamefont
  {Esry}},\ }\bibfield  {title} {\bibinfo {title} {{Hartree-Fock theory for
  Bose-Einstein condensates and the inclusion of correlation effects}},\ }\href
  {https://doi.org/10.1103/PhysRevA.55.1147} {\bibfield  {journal} {\bibinfo
  {journal} {Phys.~Rev.~A}\ }\textbf {\bibinfo {volume} {55}},\ \bibinfo
  {pages} {1147} (\bibinfo {year} {1997})}\BibitemShut {NoStop}%
\bibitem [{\citenamefont {Luo}(2009)}]{2009luo0a}%
  \BibitemOpen
  \bibfield  {author} {\bibinfo {author} {\bibfnamefont {C.}~\bibnamefont
  {Luo}},\ }\bibfield  {title} {\bibinfo {title} {{Modeling Bose-Einstein
  condensate with Gross-Pitaevskii equation}},\ }\href
  {https://web.archive.org/web/20100529054157/http://www.grinnell.edu/files/downloads/draft.pdf}
  {\bibfield  {journal} {\bibinfo  {journal} {preprint}\ } (\bibinfo {year}
  {2009})}\BibitemShut {NoStop}%
\bibitem [{\citenamefont {Bandrauk}\ and\ \citenamefont
  {Shen}(1992)}]{1992bandrauk0a}%
  \BibitemOpen
  \bibfield  {author} {\bibinfo {author} {\bibfnamefont {A.~D.}\ \bibnamefont
  {Bandrauk}}\ and\ \bibinfo {author} {\bibfnamefont {H.}~\bibnamefont
  {Shen}},\ }\bibfield  {title} {\bibinfo {title} {{Higher order exponential
  split operator method for solving time-dependent Schr{\"o}dinger
  equations}},\ }\href {https://doi.org/10.1139/v92-078} {\bibfield  {journal}
  {\bibinfo  {journal} {Can.~J.~Chem.}\ }\textbf {\bibinfo {volume} {70}},\
  \bibinfo {pages} {555} (\bibinfo {year} {1992})}\BibitemShut {NoStop}%
\bibitem [{\citenamefont {Jackson}\ \emph {et~al.}(2021)\citenamefont
  {Jackson}, \citenamefont {Corney},\ and\ \citenamefont
  {Bromley}}]{2021jackson0a}%
  \BibitemOpen
  \bibfield  {author} {\bibinfo {author} {\bibfnamefont {C.~D.}\ \bibnamefont
  {Jackson}}, \bibinfo {author} {\bibfnamefont {J.~F.}\ \bibnamefont
  {Corney}},\ and\ \bibinfo {author} {\bibfnamefont {M.~W.~J.}\ \bibnamefont
  {Bromley}},\ }\href@noop {} {\bibinfo {title} {{Benchmarking numerical
  solutions to the doubly-time-dependent Schr\"{o}dinger equation}}} (\bibinfo
  {year} {2021}),\ \bibinfo {note} {in preparation}\BibitemShut {NoStop}%
\bibitem [{\citenamefont {Dobson}(1994)}]{1994dobson0a}%
  \BibitemOpen
  \bibfield  {author} {\bibinfo {author} {\bibfnamefont {J.~F.}\ \bibnamefont
  {Dobson}},\ }\bibfield  {title} {\bibinfo {title} {{Harmonic-Potential
  Theorem: Implications for Approximate Many-Body Theories}},\ }\href
  {https://doi.org/10.1103/PhysRevLett.73.2244} {\bibfield  {journal} {\bibinfo
   {journal} {Phys.~Rev.~Lett.}\ }\textbf {\bibinfo {volume} {73}},\ \bibinfo
  {pages} {2244} (\bibinfo {year} {1994})}\BibitemShut {NoStop}%
\bibitem [{\citenamefont {Dennis}\ \emph {et~al.}(2013)\citenamefont {Dennis},
  \citenamefont {Hope},\ and\ \citenamefont {Johnsson}}]{2013dennis0a}%
  \BibitemOpen
  \bibfield  {author} {\bibinfo {author} {\bibfnamefont {G.~R.}\ \bibnamefont
  {Dennis}}, \bibinfo {author} {\bibfnamefont {J.~J.}\ \bibnamefont {Hope}},\
  and\ \bibinfo {author} {\bibfnamefont {M.~T.}\ \bibnamefont {Johnsson}},\
  }\bibfield  {title} {\bibinfo {title} {{XMDS2: Fast, scalable simulation of
  coupled stochastic partial differential equations}},\ }\href
  {https://doi.org/10.1016/j.cpc.2012.08.016} {\bibfield  {journal} {\bibinfo
  {journal} {Comp.~Phys.~Comm.}\ }\textbf {\bibinfo {volume} {184}},\ \bibinfo
  {pages} {201} (\bibinfo {year} {2013})}\BibitemShut {NoStop}%
\end{thebibliography}%

\end{document}

\title{Supplementary: Atomtronic Many-Body Transport using Husimi Driving}

\author{B.J.~{Mommers}}
\affiliation{Centre for Engineered Quantum Systems, The University of Queensland, St Lucia, Australia}
\affiliation{School of Mathematics and Physics, The University of Queensland, St Lucia, Australia}
\email[]{b.mommers@uq.edu.au}
\author{A.~{Pritchard}}
\affiliation{Centre for Engineered Quantum Systems, The University of Queensland, St Lucia, Australia}
\affiliation{School of Mathematics and Physics, The University of Queensland, St Lucia, Australia}
\author{T.A.~{Bell}}
\affiliation{Centre for Engineered Quantum Systems, The University of Queensland, St Lucia, Australia}
\affiliation{School of Mathematics and Physics, The University of Queensland, St Lucia, Australia}
\author{R.N.~{Kohn, Jr.}}
\affiliation{Space Dynamics Laboratory, Albuquerque, New Mexico 87106, USA}
\author{S.E.~{Olson}}
\affiliation{Air Force Research Laboratory, Kirtland AFB, New Mexico 87117, USA}
\author{M.~{Baker}}
\affiliation{Quantum Technologies, Cyber and Electronic Warfare Division, Defence Science and Technology Group, Brisbane, Australia}
\affiliation{School of Mathematics and Physics, The University of Queensland, St Lucia, Australia}
\affiliation{Centre for Engineered Quantum Systems, The University of Queensland, St Lucia, Australia}
\author{M.W.J.~{Bromley}}
\affiliation{School of Sciences, University of Southern Queensland, Toowoomba, Australia}
\affiliation{School of Mathematics and Physics, The University of Queensland, St Lucia, Australia}

\maketitle

\section{1. Non-linear Schr{\"o}dinger equation theory} \label{sec:theory}

In this first Supplementary section we derive the solution to the Gross-Pitaevskii equation (GPE) under Husimi driving and show that the result is independent of both the interaction term in the GPE and the initial state in the static harmonic trap. These attributes allow arbitrary harmonically-trapped states of a Bose-Einstein condensate (BEC) to be spatially driven without excitation.
The Husimi ``see-saw" potential is a type of forced harmonic oscillator which is described by a harmonic potential with linear spatial modulation,
\begin{equation}
  V(x,t) =  \frac12 m \omega_1^2 x^2 - A x f(t)
        \equiv \frac12 m \omega_1^2
        \left(x - \frac{A f(t)}{m\omega_1^2}\right)^2
              - \frac{A^2 f^2(t)}{2m\omega_1^2} \; ,
\label{eq:driving-potential}
\end{equation}
where $\omega_1$ represents the (fixed) trapping frequency of the time-independent harmonic trapping component, $f(t)$ is an arbitrary modulation function, and $A$ is the modulation amplitude.
Husimi's original solution describes the driving of a wavefunction in a harmonic
trap under the Schr{\"o}dinger equation~\cite{1953husimi0a}.
We instead consider dynamics governed by the
Gross-Pitaevskii equation (GPE), a nonlinear Schr{\"o}dinger equation (NLSE), 
\begin{equation}
i \hbar \frac{\partial}{\partial t} \psi(x,t) = \left( \! -\frac{\hbar^2}{2m} \frac{\partial^2}{\partial x^2}
                                                        + V(x,t) + g |\psi(x,t)|^2 \! \right) \! \psi(x,t) \; ,
    \label{eq:gpenlse}
\end{equation}
where $m$ is the atomic mass, the external potential
$V(x,t)$ is described by Eqn.~\ref{eq:driving-potential}.
The 1-D nonlinear interaction strength $g \propto g_{3D}$, the 3-D strength, where
$g_{3D} = 4\pi\hbar^2a_s (N-1)/m$, which depends on the
$s$-wave atom-atom scattering length ($a_s$), and the number of other atoms $(N-1)$ in the condensate~\cite{1997esry0a}.  This is then a single particle model,
where $\int |\psi(x,t)|^2 dx = 1$ is the chosen normalisation.

Following Husimi's method, we perform a coordinate transformation $x \rightarrow (x - \xi(t)) = x'$, where $\xi$ is an arbitrary function of time only. Expressing the NLSE in terms of this new coordinate gives,
\begin{equation}
i \hbar \frac{\partial}{\partial t} \psi(x'+\xi,t) = \left( i \hbar \dot{\xi} \frac{\partial}{\partial x'} - \frac{\hbar ^2}{2m} \frac{\partial ^2}{\partial {x'}^2} + \frac{m}{2} \omega_1 ^2 (x' + \xi)^2 - A (x' + \xi) f(t) + g |\psi(x'+\xi,t)|^2 \right) \psi(x'+\xi,t)~.
\end{equation}
As an ansatz, the solution is factored into the product of a phase and arbitrary function $\varphi({x'} + \xi,t)$, taking the form,
\begin{equation}
\psi(x'+\xi,t) = \exp\left( \frac{i m \dot{\xi} x'}{\hbar} \right) \varphi(x'+\xi,t)~.
\end{equation}
We insert this ansatz into the coordinate-transformed NLSE to enable terms to be separated,
\begin{equation}
  \begin{split}
i \hbar \frac{\partial}{\partial t} \varphi(x'+\xi,t) = &-\frac{\hbar ^2}{2m} \frac{\partial ^2 }{\partial {x'}^2} \varphi(x'+\xi,t) + \frac{m}{2} \omega_1 ^2 {x'}^2 \varphi(x'+\xi,t) + g |\psi(x'+\xi,t)|^2 \varphi(x'+\xi,t) \\
&\qquad + \left[ m \ddot{\xi} + m \omega_1 ^2 \xi - A f(t) \right] \ x'\ \varphi(x'+\xi,t)  \\
&\qquad + \left( - \frac{m}{2} \dot{\xi}^2 + \frac{m}{2} \omega_1 ^2 \xi ^2 -  A \xi f(t) \right) \varphi(x'+\xi,t)~.
  \end{split}
  \label{eq:transformed-NLSE}
\end{equation}
By constraining $\xi$ to satisfy the classical equation of motion for a forced oscillator $m \ddot{\xi} + m \omega_1 ^2 \xi = A f(t)$, the term in square brackets above is reduced to zero, and $\xi$ can be physically interpreted as the centre-of-mass coordinate $\langle x \rangle $.

As in Husimi's original derivation, the term within parentheses in Eqn.~\ref{eq:transformed-NLSE}
above can be reduced to zero by performing the additional factorisation,
\begin{equation}
\varphi(x' + \xi,t) = \phi(x' + \xi,t) \cdot \exp \left( \frac{i}{\hbar} \int _0 ^t \frac{1}{2} m \dot{\xi}^2 - \frac{m}{2} \omega_1 ^2 \xi ^2 + A \xi f(t) \ dt \right)~.
\end{equation}
To see the effect this factorisation has on the transformed NLSE of Eqn.~\ref{eq:transformed-NLSE}, we calculate the relevant derivatives,
\begin{align}
    \frac{\partial}{\partial t} \varphi(x' + \xi,t) &= i\hbar \frac{\partial \phi(x' + \xi,t)}{\partial t} \cdot \exp \left(\frac{i}{\hbar} \int_0^t \frac{m}{2} \dot{\xi}^2 - \frac{m}{2} \omega_1^2 \xi^2 + A \xi f(t) dt \right) \nonumber \\
 &\qquad - \left( \frac{m}{2} \dot{\xi}^2 - \frac{m}{2} \omega_1 ^2 \xi ^2 + A \xi f(t)  \right) \phi(x' + \xi,t) \cdot \exp \left( \frac{i}{\hbar} \int _0 ^t \frac{m}{2} \dot{\xi}^2 - \frac{m}{2} \omega_1 ^2 \xi ^2 + A \xi f(t) \ dt \right)~, \\
    \frac{\partial^2}{\partial x'^2} \varphi(x' + \xi,t) &= \frac{\partial^2 \phi(x' + \xi,t)}{\partial x'^2} \cdot \exp \left(\frac{i}{\hbar} \int_0^t \frac{m}{2} \dot{\xi}^2 - \frac{m}{2} \omega_1^2 \xi^2 + A \xi f(t) dt \right)~,
\end{align}
and substitute them back into the transformed NLSE. We then obtain,
\begin{align}
    i \hbar \frac{\partial \phi(x' + \xi,t)}{\partial t}
    \cdot \exp\left(\frac{i}{\hbar} \int_0^t L\ dt\right) &=
    -\frac{\hbar^2}{2m} \frac{\partial^2 \phi(x' + \xi,t)}{\partial x'^2}
    \cdot \exp\left(\frac{i}{\hbar} \int_0^t L\ dt\right)
    + \frac{m}{2} \omega_1 ^2 x^2 \phi(x'+\xi,t)
    \cdot \exp\left(\frac{i}{\hbar} \int_0^t L\ dt\right) \nonumber \\ 
&\qquad + g |\phi(x'+\xi,t)|^2 \phi(x'+\xi,t)
    \cdot \exp\left(\frac{i}{\hbar} \int_0^t L\ dt\right) \nonumber \\
&\qquad + \left[ m \ddot{\xi} + m \omega_1 ^2 \xi - A f(t) \right] \ x'\ \phi(x'+\xi,t)
    \cdot \exp\left(\frac{i}{\hbar} \int_0^t L\ dt\right) \nonumber \\
&\qquad + \left( - \frac{m}{2} \dot{\xi}^2 + \frac{m}{2} \omega_1 ^2 \xi ^2 -  A \xi  f(t) \right) \phi(x'+\xi,t)
    \cdot \exp\left(\frac{i}{\hbar} \int_0^t L\ dt\right) \nonumber \\
&\qquad + \left( \frac{m}{2} \dot{\xi}^2 - \frac{m}{2} \omega_1 ^2 \xi ^2 + A \xi f(t)  \right)\phi(x'+\xi,t)
    \cdot \exp\left(\frac{i}{\hbar} \int_0^t L\ dt\right)~,
\end{align}
where $L = \frac{m}{2} \dot{\xi}^2 - \frac{m}{2} \omega_1^2 \xi^2 + A \xi f(t)$.
The global phase factor $\exp(\frac{i}{\hbar} \int_0^t L\ dt) = \exp(\frac{i}{\hbar} Et)$ can be eliminated, and by applying the constraint on $\xi$, our result is the NLSE in terms of $\phi(x,t)$ within a static harmonic potential (changing $\partial/\partial x^{\prime} \to \partial/\partial x$),
\begin{align}
        i \hbar \frac{\partial \phi(x,t)}{\partial t} &= -\frac{\hbar^2}{2m} \frac{\partial^2 \phi(x,t)}{\partial x^2} + \frac{m}{2} \omega_1 ^2 x^2 \phi(x,t) + g |\phi(x,t)|^2 \phi(x,t)~.
    \label{eq:husimi-derivation-solution}
\end{align}
Note this result does not depend on the form of the interaction term $g|\phi(x'+\xi,t)|^2$.
To satisfy Eqn.~\ref{eq:husimi-derivation-solution}, $\phi$ must be \textit{any} solution to the NLSE for a static harmonic potential (including time-dependent solutions), which takes the form $\phi(x,t) \equiv \chi(x-\xi,t)$. This gives our final result: BEC states confined using the Husimi see-saw potential in Eqn.~\ref{eq:driving-potential} satisfy
\begin{equation}
\psi(x,t) = \chi(x-\xi,t) \cdot \exp \left( \frac{i m \dot{\xi} (x-\xi)}{\hbar} + \frac{i}{\hbar} \int _0 ^t \frac{1}{2} m \dot{\xi}^2 - \frac{m}{2} \omega_1 ^2 \xi ^2 + A \xi f(t) \ dt \right)~.
\label{eqn:husimi-GPE-solution}
\end{equation}
% Compacting the phase factor we use an equivalent short form,
% \begin{equation}
% \psi(x,t) = \chi(x-\xi,t) \hspace{0.2ex} \exp \hspace{-0.7ex} \left[ i \hspace{0.2ex} \dot{\xi} (x-\xi)/\hbar + i E \hspace{0.1ex} t/\hbar \right] ~.
% \label{eqn:husimi-soln}
% \end{equation}
This solution results in the centre-of-mass coordinate being translated according to $\xi(t)$, and the local evolution of the state about the centre-of-mass evolving according to the dynamics of the state $\chi$ in a stationary harmonic trap.

These results also hold for the full 3-D NLSE as no additional modes are excited by the Husimi driving potential.
Therefore if $\chi(x,y,z,t)$ begins in an NLSE eigenstate at $t=0$ then
the mean-field potential,
$g^\prime |\chi(x-\xi,y,z,t)|^2$,
is constant such that the dynamics all remain solely along the $x$-direction.
If it is not in an initial eigenstate then the
NLSE solution $\chi(x-\xi,y,z,t)$ will undergo dynamics
in the $x$-direction that couple energy into/out of transverse modes,
but these will remain overall unaffected by the Husimi driving.
Additionally, the independence of our theoretical solution
to inter-particle interactions indicates that
Husimi oscillations can be induced in degenerate and/or thermal ensembles of trapped atoms, whether bosonic or fermionic.

The initial motivation to investigate the above analytic solution to the NLSE was
provided by an (unpublished) NLSE computational methods report by Luo~\cite{2009luo0a},
which relied on Bandrauk and Shen's numerical work~\cite{1992bandrauk0a},
which, in turn, cites Husimi~\cite{1953husimi0a}.
Luo, however, only considered the squeezing potential,
$V(r,t) = \frac12 r^2 (1 + A \sin(\omega t))$ to explore time-dependent numerical
solutions.   Our theory here provides for an exacting test for
non-linear Schr{\"o}dinger equation computational modelling,
beyond the linear~\cite{2021jackson0a}.

\section{2. Husimi driving of many-body systems}

The main result of the previous section provides a solution to the GPE under Husimi driving, and shows that condensates described by the GPE can be prepared in arbitrary states of a harmonic trap and have their centre-of-mass driven whilst local evolution is unaffected by driving. GPE dynamics assume a homogeneous zero-temperature gas of particles in the ground state, with interactions dominated by $s$-wave scattering. This is a good approximation for very low temperature condensates. However, experimental realisations of BECs include a thermal component due to imperfect evaporation. Nevertheless, the result of the previous section can readily be generalized to an interacting many-body Hamiltonian for $N$ particles under Husimi driving. In this section we briefly show that the interaction-independence of the GPE result extends to many-body systems with arbitrary interaction potentials.

The Hamiltonian describing a many-body system, within the external potential Eqn.~\ref{eq:driving-potential}, takes the (1-D) form
\begin{equation}
        H = \sum_{j=1}^N \left(\frac{-\hbar^2}{2m} \frac{\partial ^2}{\partial x_j ^2} + \frac{1}{2} m \omega_1 ^2 x_j ^2 - A x_j f(t)  + \sum _{k\neq j} ^N V_{\text{int}}(x_j,x_k) \right)~,
\end{equation}
where $V_{\text{int}}$ is an arbitrary interaction potential describing pairwise interparticle interactions
(and not double counting).
Performing the change of coordinates $x_j \rightarrow (x_j - \xi(t)) = x_j^\prime$ we obtain the many-body Schr{\"o}dinger equation,
\begin{equation}
\label{eqn:manybodytdse}
 \begin{split}
        \sum_{j=1}^N  i \hbar \frac{\partial}{\partial t} \psi (x_j ^{\prime} + \xi, t)
      = \sum_{j=1}^N \bigg( i \hbar \dot{\xi} \frac{\partial}{\partial x_j ^{\prime}}
                            - \frac{\hbar^2}{2m} \frac{\partial ^2}{\partial x_j ^{\prime 2}}
                            + \frac{1}{2} m \omega_1 ^2 (x^{\prime} _j + \xi ) ^2
                            & - A (x^{\prime}_j + \xi ) f(t)  + \xi, t) \\
           & + \sum _{k\neq j} ^N V_{\text{int}}(x^{\prime}_j+\xi,x_k^{\prime}+\xi)\bigg) \psi (x_j ^{\prime} + \xi, t) ~.
 \end{split}
\end{equation}
The form of this transformed Hamiltonian shows that each particle in an arbitrary many-body state under Husimi driving evolves locally
(with respect to the centre-of-mass motion) as if in a static harmonic trap, whilst being translated according to the classical forced oscillator solution.
For an experimental BEC, this results in co-driving of both the condensate and any thermal cloud,
indicating that driving will not induce additional heating, atom loss, or other excitations.

The Husimi-based solution to Eqn.~\ref{eqn:manybodytdse} was also derived by Dobson
in his development of the many-body Harmonic Potential Theorem (HPT)~\cite{1994dobson0a}. 
The HPT further generalises the generalized Kohn theorem (GKT) to describe the dynamics of many-body
interacting systems in an eigenstate of a harmonic scalar potential~\cite{1994dobson0a}.
By performing the same coordinate transformation to that used by Husimi, the solution to
the GKT Hamiltonian for interacting gases in a harmonic potential with a uniform linear field
can be separated into the relative motion of the centre-of-mass and the
stationary many-body eigenstate --- the main result of the HPT.
Our present work is a further generalisation of the HPT beyond many-body eigenstates, as we have shown in the previous section that arbitrary many-body states are preserved in the co-moving centre-of-mass frame.

\section{3. Phenomenological damping of Husimi trajectories}

The lack of damping in the all-magnetic atom chip experimental trajectories, as shown in Fig.~4 of the main text via the fitted $\gamma = 0.00 \pm 0.02$~s$^{-1}$, indicates that the finite temperature of a condensate in the lab does not result in damping by itself.
This is in agreement with the many-body theoretical result (above) where Husimi driving dynamics apply equally to condensed and thermal atoms.

In contrast, our optical dipole experimental investigation was initially unable to observe Husimi driving due to a poor optical pinning beam spatial distribution.
After several adjustments to the beam, the first signatures of Husimi oscillations were observed -- though damping still occurred over short time-scales. 
At this point a damping term was added to the analytical expression for the centre of mass trajectory (Eqn.~6 in the main text) in order to help quantify the effects of subsequent experimental changes on trajectory damping.
As shown in Figs.~2 and 3 of the main text, damping was improved enough to observe Husimi driving, but remains significant, and from our experimentation appears to be independent of driving frequency but strongly dependent on temperature.
The temperature-dependence was experimentally tested by varying the evaporation process to prepare clouds with different condensate fractions and therefore temperatures.

\begin{figure}
    \centering
    \includegraphics[width=\linewidth]{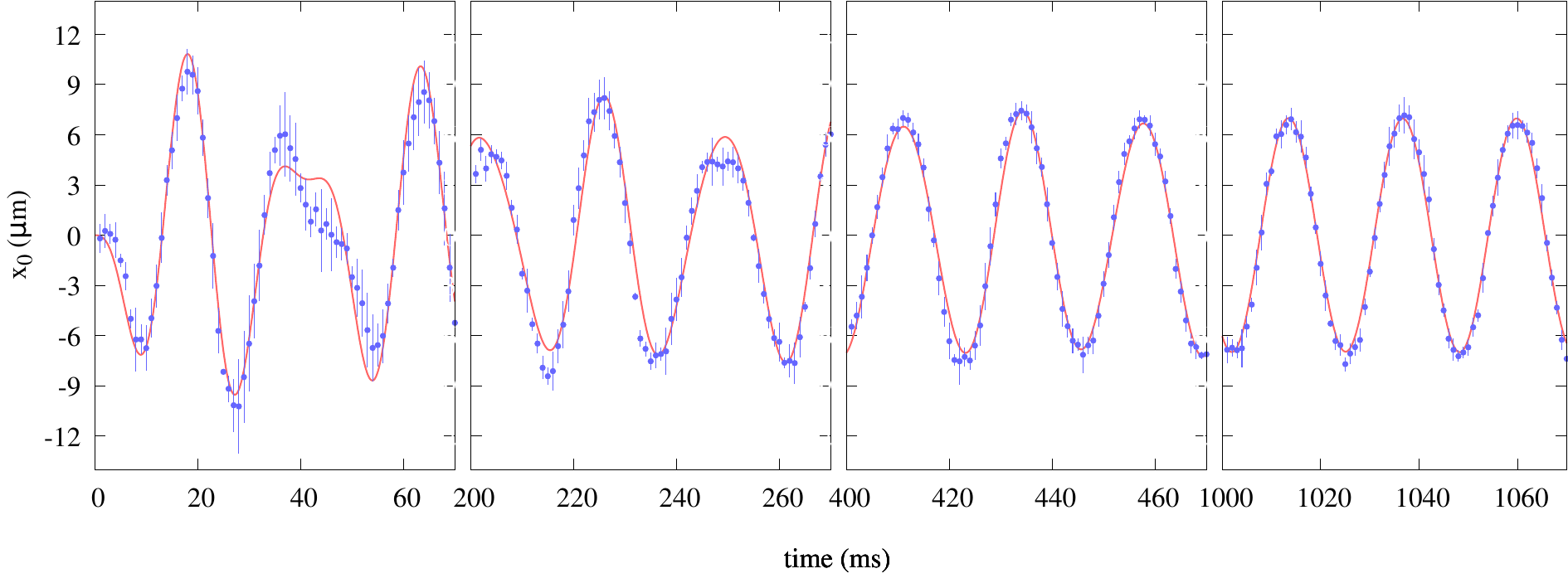}
    \caption{
    (color online) Optical dipole experiment under Husimi driving
with trap frequency of $\omega_1 \approx 2\pi \cdot 61\pm3$~Hz
and periodic driving frequency set to $\omega_2 = 2\pi \cdot 43.13$~Hz to give an initial aperiodic trajectory that then shows damping of the trajectory towards the driving frequency.
The experimental configuration in these experiments is slightly different to those in Fig.~2 of the main text, with a stronger magnetic field driving larger spatial oscillations.
The data was taken over $70$~ms windows after driving for $\{0,200,400,1000\}$~ms to assess the long-time effects of damping on trajectories.
Only shown here are the centre-of-mass measurements in the $x$-direction.
Each data point is the average of three runs, with one standard deviation indicated by the error bars.
The solid (red) line shows the fit over the entire dataset (to Eqn.~6 in the main text) with parameters:
$\omega_1 = 2\pi \cdot 67.43 \pm 0.06$~Hz,
$\omega_2 = 2\pi \cdot 43.17 \pm 0.01$~Hz,
$b_x = 2.30 \pm 0.03$~G/cm,
$\gamma = 5.4 \pm 0.4$~s$^{-1}$.
}
    \label{fig:long-time-data}
\end{figure}

We decided to perform additional optical dipole experiments,
going out beyond $1$~second of driving time.
Supplemental Fig.~\ref{fig:long-time-data} shows very clearly the effect of damping over such long times; the centre-of-mass trajectory begins by following the Husimi solution until it evolves to follow the driving frequency at long times.
(c.f. Fig.~2 of the main text).

To understand this further we have explored the optical dipole
trapping system numerically, and show here that anharmonic noise/spatial distortions across the Gaussian potential reproduce the order of magnitude of the experimentally observed damping behaviour.
Two effects were initially suspected: (1) imperfect beam propagation through the experimental optical system, and (2) the Gaussian pinning beam intensity profile itself.
These are the two most conspicuous sources of anharmonicity explored numerically.

Several numerical potentials based on the experimental parameters were constructed to explore various damping contributions in isolation.
An experimentally representative distorted Gaussian potential, $V_\text{Experiment}$, was firstly constructed using the experimentally measured optical intensity profile; the intensity distribution was measured using our standard $M=6.38$ vertical imaging system and Prosilica EC380 camera.
A purely Gaussian potential $V_{\mathrm{Gaussian}}$
was constructed by taking the numerical regression with $V_{\mathrm{Experiment}}$, viz.
\begin{equation}
    V_{\mathrm{Gaussian}}(x,y) = -V_0 \exp\left(-\frac{2\left(x-x_0\right)^2}{w_x^2}\right)  \exp\left(-\frac{2\left(y-y_0\right)^2}{w_y^2}\right) \; ,
\end{equation}
where $V_0 = 3480$~nK and the widths, $w$, chosen to match the experiment (i.e. to match the pinning beam widths given in the main text, and the frequency as determined by parametric heating, $61$~Hz).

This allows us to approximate the residual corrugations as
$V_{\mathrm{Noise}} = V_{\mathrm{Experiment}} - V_{\mathrm{Gaussian}}$, which contains amplitudes of order $\sim 0.05 V_0$.
Furthermore, an idealized harmonic potential $V_{\mathrm{Harmonic}}$ based on the quadratic fit to $V_{\mathrm{Gaussian}}$ was constructed to enable the gradual introduction of the higher-order anharmonic distortions (and provided an initial means to evaluate the adequate convergence of our numerical results against the exact analytic results).
Finally, a corrugated harmonic potential was introduced
$V_{\mathrm{Corrugated}} = V_{\mathrm{Harmonic}} + V_{\mathrm{Noise}}$.

Two-dimensional Gross-Pitaevskii simulations were performed using the XMDS2 numerical integration package~\cite{2013dennis0a}. These simulations adopted a $1024 \times 1024$ point spatial lattice which spanned $\pm 34.304$~$\mu$m along the $x$ and $y$-dimensions. Dynamical processes along the third $z$-dimension were neglected given that comparatively tight confinement ($\omega = 2\pi \cdot 140$~Hz) was provided by the sheet potential, rather than the pinning beam ($\omega_1 \approx 2\pi \cdot 61$~Hz).
Effectively zero-temperature condensates with chemical potential $\mu = 80$~nK and thus $N = 2.2 \times 10^5$ atoms were prepared using
imaginary time methods, which gives the initial ($t=0$) wavefunction.
The Husimi driving potential was chosen to have field gradient $b_x = 1.04$~G/cm, with frequency $\omega_2=43.13$~Hz (giving dynamics similar to Fig.~2(a-c) in the main text).
Our Husimi simulations used the fourth-order Runge-Kutta (RK4) algorithm and $1$~$\mu$s fixed integration time-steps.

The Husimi damping strength associated with the distortions introduced by the optical system or associated with the Gaussian beam profile were investigated using the dimensionless parameters $\delta$ and $\beta$, through the continuously variable potentials,
\begin{align}
    V(\delta) = (1-\delta) V_{\text{Harmonic}} + \delta \, V_{\text{Corrugated}} \quad , \quad
    V(\beta) &= (1-\beta) V_{\text{Harmonic}} + \beta V_{\text{Gaussian}} \;.
    \label{eq:xmds-potential}
\end{align}
Having run each $\delta$ or $\beta$ simulation, we fit to the trajectories, and extract the estimated
amount of damping.

Our quantitative damping results, based on the optical dipole experiment, are compiled in Supplemental Fig.~\ref{fig:damping-anharmonicity-plot}.
Firstly, we note that the simple phenomenological model introduced in Eqn.~6 in the main text appears to well describe both systems as the anharmonicities are introduced.
The effect of the noise through $\delta$ is an order of magnitude larger than the inclusion of the Gaussian anharmonicity ($x^3$, $x^4$ etc) through $\beta$.
This suggests that it is the laser corrugation noise that dominates the damping of Husimi centre-of-mass oscillations in our optical dipole experiment.
We observe the corrugation noise induced damping increase up towards $3$~Hz which is consistent with the experimental values (see Fig.~3 of the main text).
Pleasingly, the calculations show that only a small damping is
introduced though the inclusion of the Gaussian anharmonicity, which suggests that our Gaussian laser potential is harmonic enough to observe Husimi driving --- as we did in the optical dipole experiments.
Also, $x^3$ and $x^4$ terms can be easily engineered in the atom chip experiment, but this was not done here.
\begin{figure}[b]
    \centering
    \includegraphics[width=0.60\linewidth]{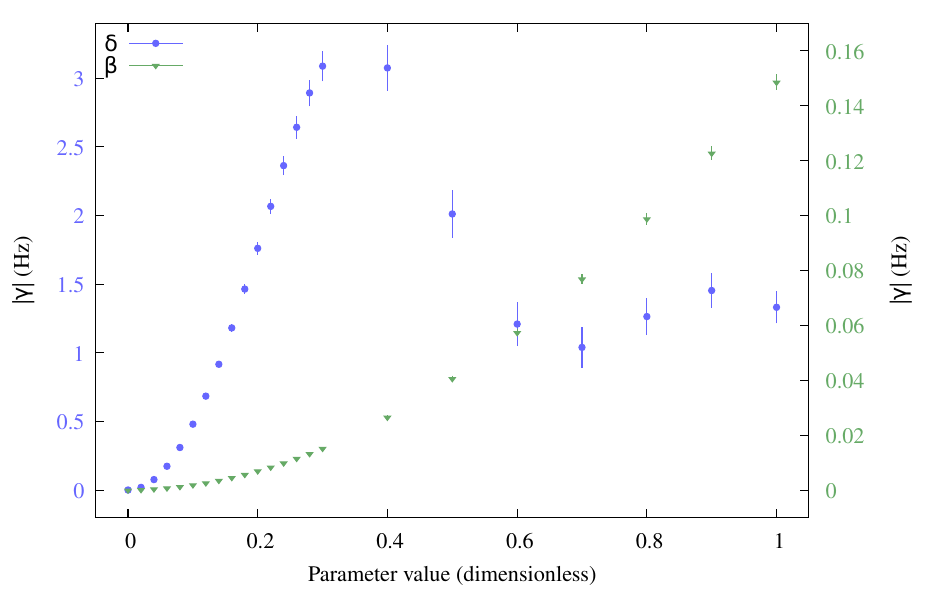}
    \caption{(color online) Simulation of the optical dipole experiment exploring the effect of trap
                     anharmonicity on damping of simulated trajectories using two different sets of potentials
                     as per Eqn.~\ref{eq:xmds-potential}.
     Parameter $\delta$ (blue circles, left vertical axis) interpolates between a smooth harmonic trap at $\delta=0$ and a corrugated harmonic trap with experimentally-observed deformation profile and amplitude at $\delta=1$; distortions with amplitudes up to of order $\sim 0.05 V_0$ are introduced for $\delta=1$,
     where $V_0 = 3480$~nK represents the Gaussian potential depth.
     Parameter $\beta$ (green triangles, right vertical axis) interpolates between an ideal harmonic trap at $\beta=0$ and an ideal Gaussian trap at $\beta=1$.
    Error bars indicate uncertainty in the best fit to simulated trajectories.
    Note the use of two different scales for the $y$-axes.
    For further calculation details see the corresponding text.
    }
    \label{fig:damping-anharmonicity-plot}
\end{figure}

Furthermore, in the simulations for $\delta > 0.3$ we observe a breakdown in the simple behaviour - the damping rate reduces whilst we observe increasing centre-of-mass oscillations in the direction transverse to the drive, induced by the measured noise profile present in the experimental apparatus.
This regime requires further investigation to understand, but was initially suspected to be numerical of origin as more highly-excited modes are initially created in the wavefunction at $t=0$, and then even more are generated by the time-dependent scattering of the wavefunction as it is driven through the static, yet noisy, corrugated potential.
The grid effectively places an upper limit on the excitations that can be generated in the calculation, however, running simulations
with $2048 \times 2048$ grid points and timesteps of $316$~ns and $100$~ns does not change these $\delta > 0.3$ results.
We speculate that the temperature-dependence of damping
(as shown in Fig.~3 of the main text) is due to the coupling of such collective excitations to the thermal cloud which is co-oscillating with the condensate through the corrugated potential in the experiment.
This would require modelling beyond the Gross-Pitaevskii Equation (Eqn.~\ref{eq:gpenlse}).

\section{4. Husimi trajectories for trap frequency measurement}

Whilst Husimi trajectory fitting allows for determination of trap frequency, its utility in practice requires accurate and precise measurement with a minimal number of data points.
In this section we consider a method for making optimal use of the analytic solution for the centre-of-mass trajectory when determining trap frequency.
The variation in the trajectory $\xi(t)$ as trap frequency changes is given by the derivative,
\begin{equation}
    \frac{\partial \xi}{\partial \omega_1} = \frac{A}{m(\omega_1^2 - \omega_2^2)} \left[ 
    \frac{1}{\omega_1} 
    \left(
    \frac{\omega_2}{\omega_1} \sin(\omega_1 t) - \omega_2 t \cos(\omega_1 t)
    \right)
    - \frac{2}{(\omega_1^2 - \omega_2^2)}
    \left(
    \sin(\omega_2 t) - \frac{\omega_2}{\omega_1} \sin(\omega_1 t)
    \right)
    \right] ~ .
\end{equation}
The $-\omega_2 t \cos(\omega_1 t)$ term in this expression indicates that deviations in the trap frequency produce more distinguishable trajectories at later times, therefore late-time measurements should allow for more precise frequency fitting.
However, condensate lifetime and imperfections in the experimental implementation of Husimi driving will constrain the maximum duration for which the state is preserved.
Therefore we propose a method for making optimal use of the known analytic solution for improving trap frequency measurement precision.
\par
First, estimate the trap frequency based on known experimental parameters.
Then compare the analytic trajectory at this frequency to trajectories for nearby trap frequencies,
such as in Supplemental Fig.~\ref{fig:trap-deviation}.
This allows identification of regions in which features are qualitatively distinguishable for small variations in trap frequency.
Trajectory data should be focused around these features to constrain the fit.
For similar parameter values as used in the optical dipole trap experiments in the main text,
Supplemental Fig.~\ref{fig:trap-deviation} indicates that sub-Hz precision can be achieved
by focusing data collection around specific regions of the trajectory,
before the trajectories diverge with time.
\begin{figure}[b]
    \centering
    \includegraphics[width=0.8\linewidth]{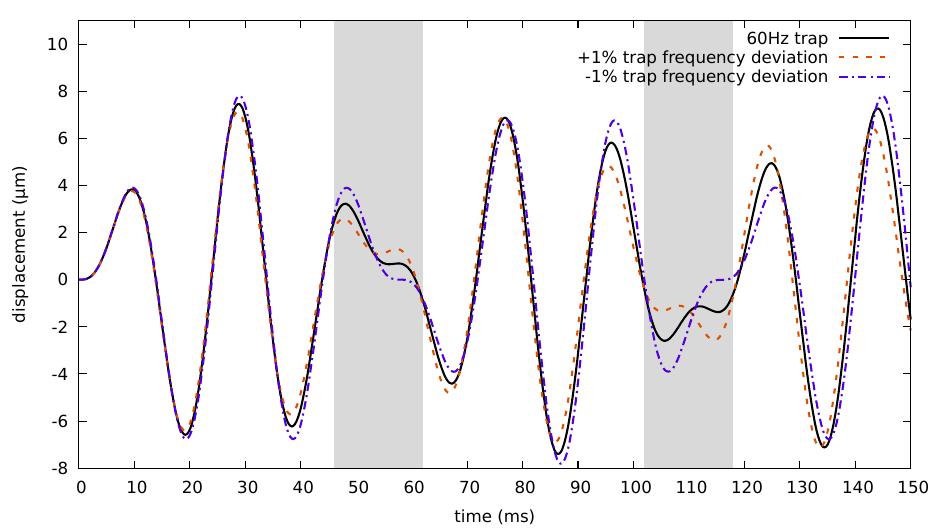}
    \caption{(color online) Analytic trajectories for Husimi driving at fixed driving frequency
    $\omega_2 = 2\pi \cdot 43.134$~Hz with driving amplitude $A = 1.0$~G/cm
    and varying trap frequencies $\omega_1 = 2\pi \times 60$~Hz with a $1\%$ deviation.
    Features around $t\approx 55$~ms and $115$~ms  (indicated by grey shaded regions) could be used to increase fit precision due to the higher distinguishability of features in the trajectory.}
    \label{fig:trap-deviation}
\end{figure}

%%%%%%%%%%%%%%%%%%%%%%%%%%%%%%%%%%%%%%%%%%%%%%%%%%%%%%%%%%%%%%%%%%%%%%%
\newpage

\section{5. Additional Experimental Data}

We wrap up this supplemental paper with an additional experiment that was performed in the optical dipole experiment, and then present two sets of original datapoints for the atom chip experiments, and a final extended experimental atom chip dataset.

A single set of near-resonant driving runs with the optical dipole experiment is shown in Supplemental Fig.~\ref{fig:UQExpData2}.
There is a small amount of cross-coupling of the direction of Husimi driving into the transverse ($y$) direction.
Driving close to the trap frequency increases the displacement at turning points in the trajectory and ensures the magnetic field gradient is flat at those turning points, creating ideal conditions to rapidly move the centre of the harmonic trap to the new location to enable the transport and relocation of a cloud of atoms.
In the current experimental setup we were unable to rapidly move the laser (which could simply be achieved with an Acousto-Optic Modulator Device or a Spatial Light Modulator), and instead we demonstrated quantum transport in the configurable atom chip device (see Fig.~5 in the main text).
\begin{figure}[h]
\includegraphics[width=0.6\linewidth]{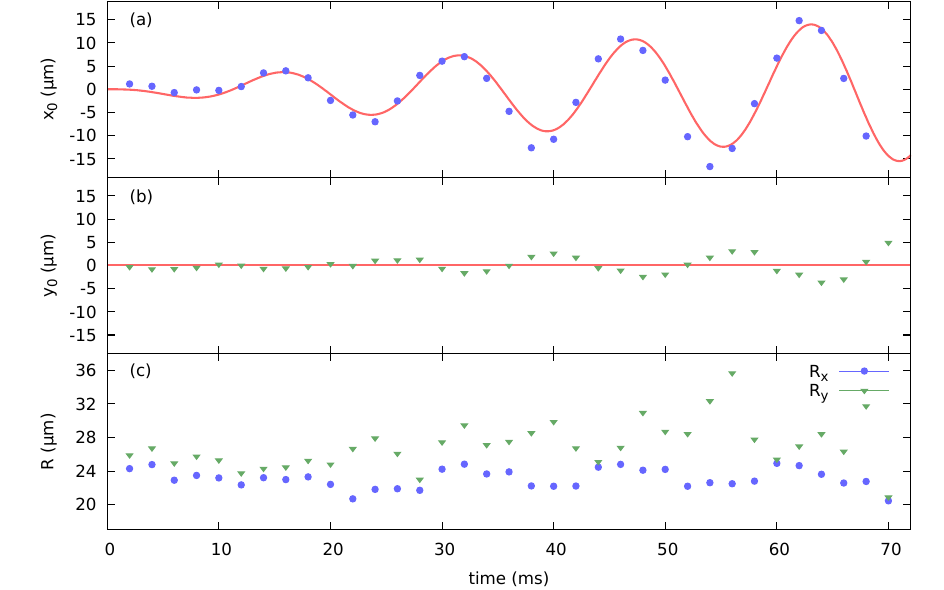}
\caption[Standard Husimi]{\label{fig:exp02-uqfast}
(color online)
Optical dipole experiment under near-resonant Husimi driving
with trap frequency of $\omega_1 \approx 2\pi \cdot 61\pm3$ Hz
and periodic driving frequency set to $\omega_2 = 2\pi \cdot 64$~Hz
to give a near-periodic trajectory.
Panels (a),(b) show centre-of-mass measurements in the $x$- and $y$-direction, respectively.
Panel (c) shows the cloud radius ($R$) measured in the $x$- and $y$-directions.
Each data point is the average of three runs, with one standard deviation indicated by
the error bars.
The solid line in (a) shows the fit to Eqn.~6 of the main text
with parameters:
$\omega_1 = 2\pi \cdot 61.9 \pm 4.8$~Hz,
$\omega_2 = 2\pi \cdot 64.9 \pm 6.7$~Hz,
$b_x = 0.566 \pm 0.041$~G/cm, $\gamma = 5.2 \pm 18$~s$^{-1}$.
}
\label{fig:UQExpData2}
\end{figure}

The four data series taken in the atom chip experiment to obtain the
aperiodic Husimi trajectories for $\omega_2/\omega_1 \approx 1/\sqrt{2}$ driving (see Fig.~4 of the main text) are given in Supplemental Fig.~\ref{fig:supp-afrl-sqrt-data}.
Strong agreement in the trajectories and only gradual long-term broadening of the cloud are observed.
A small number of outliers can be identified by the centre-of-mass position in the vertical $z$-direction.
A gradual broadening of the cloud over long times (many driving periods) is present in some of the datasets used, which is captured by the averaged data presented in Fig.~4 of the main text.
\begin{figure}[h]
\centering
\includegraphics[width=0.6\linewidth]{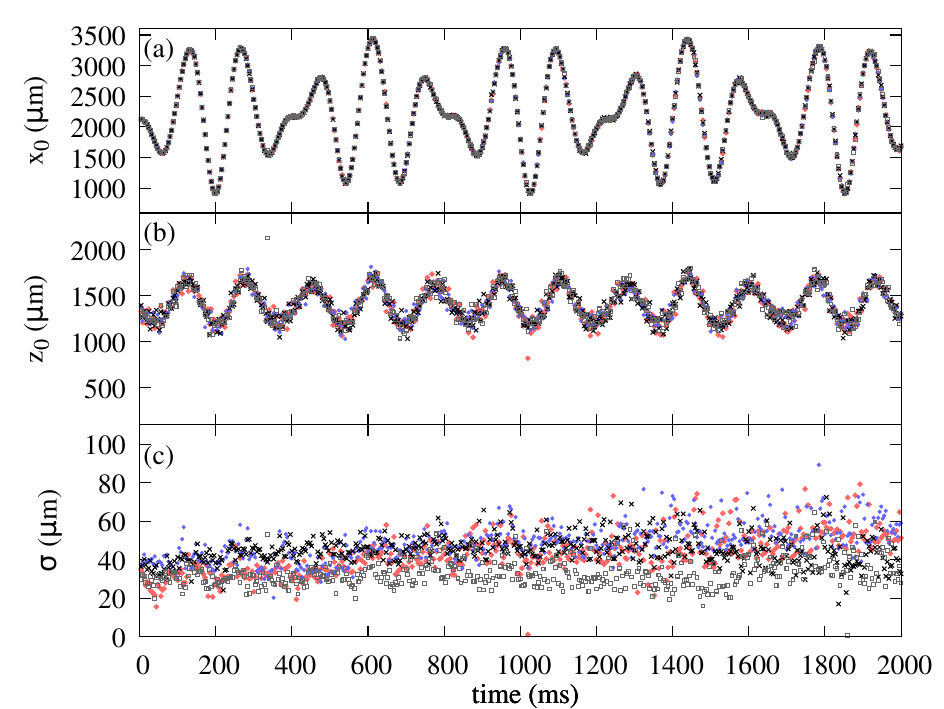}
\caption{(color online) Atom chip experiment under Husimi driving
with trap frequency of $\omega_1 \approx 2\pi \cdot 8.545\pm0.003$ Hz
and periodic driving frequency set to $\omega_2 = 2\pi \cdot 6.042$~Hz
to give an aperiodic trajectory.
Panels (a),(b) show centre-of-mass measurements in the $x$- and $z$-direction, respectively.
Panel (c) shows the cloud width measured in the $x$- and $z$-directions.
Note that the panel (b) scale is different to panel (a).
The averaged data from each of these four runs appears in Fig.~4 of the main text.
}
\label{fig:supp-afrl-sqrt-data}
\end{figure}

% \newpage

The four data series taken in the atom chip experiment to obtain the
periodic Husimi trajectories for $\omega_2/\omega_1 \approx 1$
driving (see Fig.~5 of the main text) are given in Supplemental Fig.~\ref{fig:supp-afrl-transport-data}.
Four data series were averaged for the resonant driving shown, and three data series were averaged for each of the transport protocol demonstrations.
The consistency of measured values in  Fig.~\ref{fig:supp-afrl-transport-data}(a)
indicates the robustness of the driving and transport scheme.
No broadening of the cloud width is observed for the runs in which the cloud was caught at a turning point in the trajectory to achieve
transport.

Finally, an extended time-series of the resonantly-driven atom chip dataset is provided in Supplemental Fig.~\ref{fig:supp-afrl-resonant-data}, which includes and extends the data in Supplemental Fig.~\ref{fig:supp-afrl-transport-data},
(and thus also includes and extends beyond the 'Res' data shown in Fig.~5 of the main text).
Some outlying points in the $x$-trajectory, in addition to substantial broadening of the cloud after multiple resonant driving periods are observed in some of the datasets.
Impressively, the resonant driving in the atom chip is able to extend to transport over distances of over $4$~mm, which is 100 times larger than the cloud width, and beyond which the limit of the atom chip potential control lies.

\bibliography{atom-optics-qld}

\newpage

\begin{figure}[t]
\centering
\includegraphics[width=0.6\linewidth]{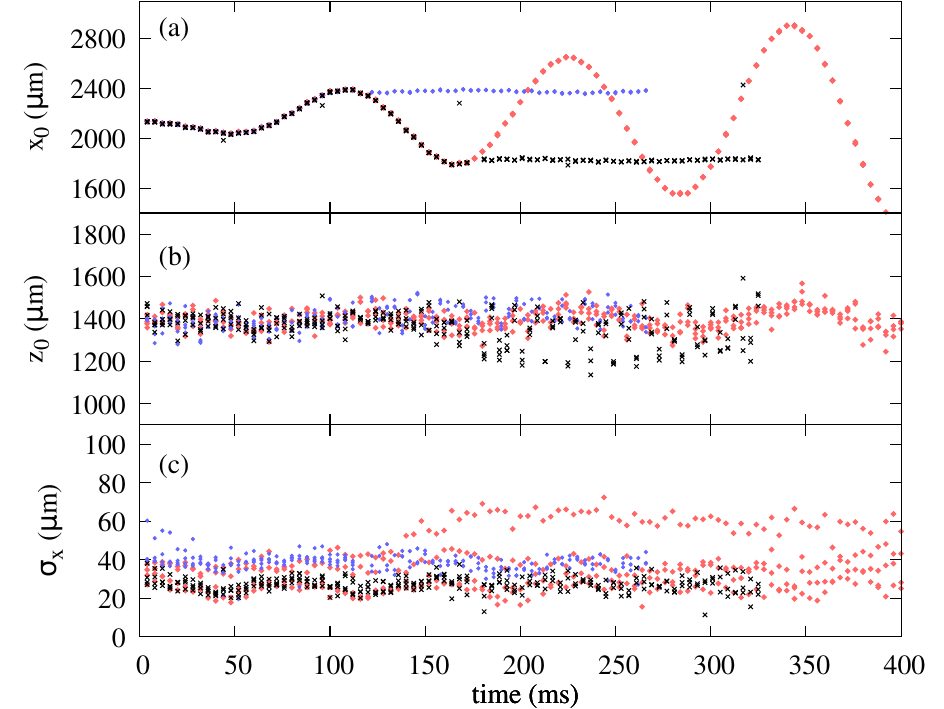}
\caption{(color online)
Atom chip experiment for Husimi driven transport
with trap frequency of $\omega_1 \approx 2\pi \cdot 8.545\pm0.003$ Hz
and resonant driving frequency set to $\omega_2 = 2\pi \cdot 8.545$~Hz.
Three experiments are shown --- red data points are the four resonant driving data runs
(see Eqn.~5 in the main text),  whilst the blue and black are
the two transport experiments that catch the atoms at the turning points, each consisting of three data runs.
Panels (a),(b) show centre-of-mass measurements in the $x$- and $z$-direction, respectively.
Panel (c) shows the cloud width ($\sigma_x$) measured in
the $x$-direction, with marker types and colours matching (a) and (b).
Note that the panel (b) scale is different to panel (a).
The averaged data from each of these four or three runs appears as `Res', `C1', and `C2'
in Fig.~5 of the main text.}
\label{fig:supp-afrl-transport-data}
\end{figure}

\begin{figure}[b]
\centering
\includegraphics[width=0.6\linewidth]{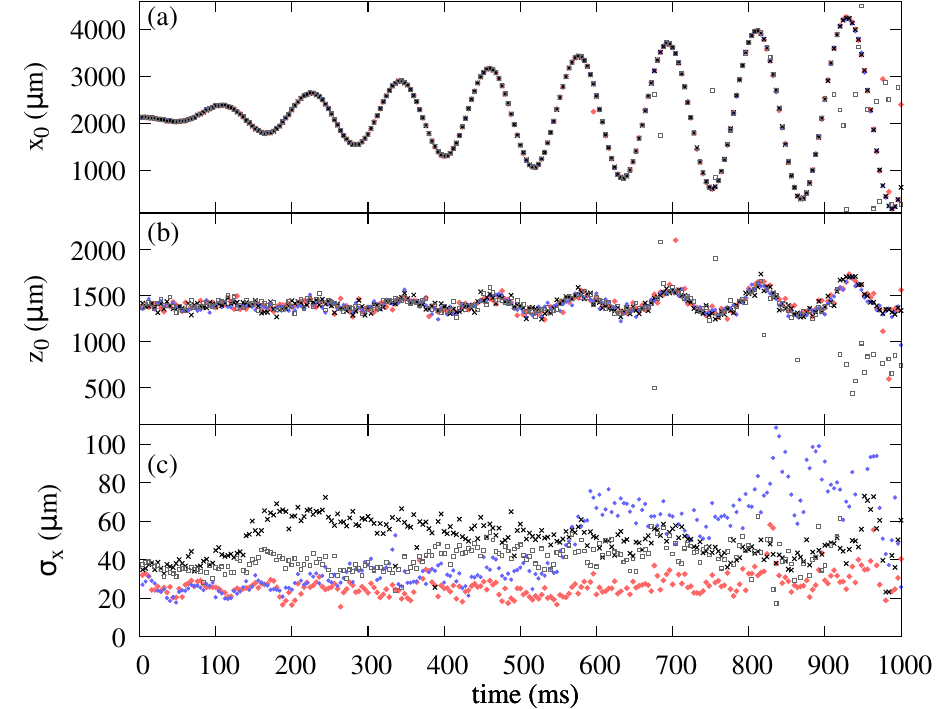}
\caption{(color online)
Atom chip experiment for Husimi driven transport
with trap frequency of $\omega_1 \approx 2\pi \cdot 8.545\pm0.003$ Hz
and resonant driving frequency set to $\omega_2 = 2\pi \cdot 8.545$~Hz.
The data points in different colours are the four resonant driving data runs (see Eqn.~5 in the main text).
Panels (a),(b) show centre-of-mass measurements in the $x$- and $z$-direction, respectively.
Panel (c) shows the cloud width ($\sigma_x$) measured in
the $x$-direction, with marker types and colours matching (a) and (b).
Note that the panel (b) scale is different to panel (a).
The averaged data from each of these four runs appears as `Res' in Fig.~5 of the main text (and unaveraged in Supplemental Fig.~\ref{fig:supp-afrl-transport-data}), however, here the experiment times were pushed out to $1$~s.
}
\label{fig:supp-afrl-resonant-data}
\end{figure}